\documentclass[10pt,a4paper,notitlepage]{article}

\usepackage{amsfonts}
\usepackage{amsmath}
\usepackage{amssymb}
\usepackage{amsthm}
\usepackage{color}
\usepackage[T1]{fontenc}
\usepackage[utf8]{inputenc}
\usepackage{graphicx}

\usepackage{mathrsfs}
\usepackage{multirow}
\usepackage[authoryear]{natbib}
\usepackage[format=hang]{subcaption}
\usepackage{times}
\usepackage{upgreek}
\usepackage{bm}
\usepackage{bbm}
\usepackage{math}			
\usepackage{mathtools}

\AtBeginDocument{} 

\usepackage{pgf}
\usepackage{tikz}
\usepackage{pgfplots}
\pgfplotsset{compat=newest}
\usepackage{pgfplotstable}
\usepgfplotslibrary{groupplots}
\usepackage{color}
\usetikzlibrary{calc,backgrounds}  
\usetikzlibrary{arrows.meta}

\definecolor{rwth1}{RGB}{0,84,159}      
\definecolor{rwth2}{RGB}{142,186,229}   
\definecolor{rwth3}{RGB}{0,97,101}      
\definecolor{rwth4}{RGB}{0,152,161}     
\definecolor{rwth5}{RGB}{87,171,39}     
\definecolor{rwth6}{RGB}{189,205,0}     
\definecolor{rwth7}{RGB}{255,237,0}     
\definecolor{rwth8}{RGB}{246,168,0}     
\definecolor{rwth9}{RGB}{227,0,102}     
\definecolor{rwth10}{RGB}{204,7,30}     
\definecolor{rwth11}{RGB}{161,16,53}    
\definecolor{rwth12}{RGB}{97,33,88}     
\definecolor{rwth13}{RGB}{122,111,172}  

\usepackage{hyperref}
\usepackage{cleveref}
\hypersetup{colorlinks,citecolor=rwth13,linkcolor=rwth13,urlcolor=rwth13}

\usepackage[left=22mm,
  right=22mm,
  top=21mm,
  bottom=29mm]{geometry}

\usepackage{booktabs} 

\usepackage{import}

\usepackage[ruled,vlined,noend]{algorithm2e}
\SetKwInput{KwInput}{Input}      
\SetKwInput{KwOutput}{Output}
\SetKwInput{kwInit}{Initialize} 
\SetKwInput{kwOffline}{Offline} 
\SetKwInput{kwOnline}{Online}

\crefname{algocf}{alg.}{algs.}
\Crefname{algocf}{Algorithm}{Algorithms} 

\crefname{equation}{eq.}{eqs.}
\Crefname{equation}{Eq.}{Eqs.} 
\crefname{figure}{fig.}{figs.}
\Crefname{figure}{Fig.}{Figs.}

\usepackage{todonotes}

\date{}

\graphicspath{{./images/}}

\begin{document}

\author{\large{Stephan Ritzert${}^{\,a}$, Jannick Kehls${}^{\,a}$,  Erik Faust${}^{\,a}$ }\\ \large{ Hagen Holthusen${}^{\,b}$}, \large{ Stefanie Reese${}^{\,a,c}$, Tim Brepols${}^{\,a}$ }\\[0.5cm] 
\hspace*{-0.1cm}
\normalsize{\em ${}^{a}$Institute of Applied Mechanics, RWTH Aachen
  University,}
\normalsize{\em Mies-van-der-Rohe-Str.\ 1, 52074 Aachen, Germany}
\\
\normalsize{\em ${}^{b}$Institute of Applied Mechanics, University of Erlangen-Nuremberg,}
\normalsize{\em Egerlandstraße\ 5, 91058 Erlangen, Germany}\\
\normalsize{\em ${}^{c}$University of Siegen, 57076 Siegen, Germany}\\
\normalsize{\em \{stephan.ritzert, jannick.kehls, erik.faust, stefanie.reese, tim.brepols\}@ifam.rwth-aachen.de}\\
\normalsize{\em hagen.holthusen@fau.de}\\[0.25cm]
}


\title{\LARGE Component-wise hyperreduction for nonlinear solid mechanics problems}
\maketitle

\small
{\bf Abstract.} 

Hyperreduced nonlinear solid-mechanics components can be
generated offline and reused as transferable building blocks across different assemblies, boundary
conditions, meshes, material parameters, and constitutive models.
We use proper orthogonal decomposition (POD) and energy conserving sampling and weighting (ECSW) on the component level and connect the substructures by mortar mesh tying. 
The  POD modes and the ECSW weights and elements are computed offline from simulations of single components and the finite rigid body motions are treated by 12 additional rigid body modes per substructure.  
The numerical examples demonstrate errors below 1 \% for large quasi-static assemblies while evaluating less than 10 \% of the elements. 
The same component bases and ECSW elements are successfully reused for finite-strain viscoelastic
dynamics, although they were trained only on elastic Neo-Hookean component simulations. 
These results indicate that component-wise hyperreduction can provide reusable reduced building blocks for modular nonlinear solid-mechanics simulations.

\vspace*{0.3cm}

\normalsize


\section{Introduction}

In nonlinear solid mechanics, parametric projection based model order reduction (MOR) methods are widely used. 
Typical applications are computational homogenization, e.g. \citep{fritzen2015nonlinear,radermacher2016displacement,zahr2017multilevel,wulfinghoff2025e3c,scheunemann2026manifold,wulfinghoff2026computational}, nonlinear vibrations, e.g. \citep{rutzmoser_model_2018,touze2021model,saccani2026accelerating} and also history-dependent material behavior involving viscoelasticity, plasticity and damage, e.g. \citep{ryckelynck2009hyper,ghavamian2017pod, kehlsMultifieldDecomposedHyperreduced2026}. 

For nonlinear problems, a common method is the proper orthogonal decompostion (POD) (\cite{chatterjee2000introduction}). 
In POD, full-order solutions for samples of input parameters are first computed in an offline stage.
These so-called snapshots are then used to compute a projection matrix, using which the FOM can be projected onto a low-dimensional subspace. 
In the online stage, the reduced order model (ROM) can be solved for new input parameters.
Because of the reduced dimension, the solution of the equation system is much faster than the FOM, but the assembly of the equation system still depends on the original size of the problem.
To accelerate the assembly, different hyperreduction methods were developed in the past two decades.  
The main idea of hyperreduction methods is to approximate the nonlinear terms in the reduced system of equations by evaluating only a subset of the elements or integration points.
There are two main families of hyperreduction method,
the  approximate-then-project  and the project-then-approximate hyperreduction methods. 
In approximate-then-project methods, the nonlinear term is first approximated using sparse reconstruction and then projected onto the reduced space. 
The most popular method of this family is the discrete empirical interpolation method (DEIM) developed by \citet{chaturantabut_nonlinear_2010} and its unassembled version (UDEIM) developed by \citet{tiso2013discrete}.
The full nonlinear term is approximated by a linear combination of modes that are computed from the snapshots of the nonlinear term. 
The coefficients are obtained by evaluating the nonlinear term in a subset of the elements or integration points. 
In solid mechanics, DEIM has been used for example in \citet{radermacher_pod-based_2016,bonomi2017matrix, ghavamian2017pod,kehls2026hyper}. 
In the project-then-approximate methods, the projected nonlinear terms of the reduced set of elements or integration points are weighted to approximate the full nonlinear term.
The first method that can be classed as part of this family was developed in \citet{an2008optimizing}.   
In \citet{farhat_structure-preserving_2015}, also the projected element quantities are weighted to approximate the full projected nonlinear term. 
The difference to \citet{an2008optimizing} is that the weights are computed by a greedy solution of a sparse non-negative least-squares problem.
The empirical cubature method (ECM) developed in \citet{hernandez_dimensional_2017} and its continuous version \citet{hernandez_cecm_2024} evaluates and weights the nonlinear terms in a subset of integration points. 
In both methods, the weights are computed by a greedy solution of a sparse non-negative least-squares problem. 
The difference is, that in ECSW the least‑squares problem is formulated using element contributions to the projected internal forces, whereas ECM approximates the full quadrature by selecting and weighting individual integration point contributions.
\citet{yano2019lp} developed a method where the weights are computed by solving a linear programming problem.
For nonlinear computational homogenization problems, recently, two new methods were developed.
\citet{wulfinghoff2025e3c,wulfinghoff2026computational} developed a method where statistical representatives of the strain mode distribution are used instead of a subset of the integration points. 
This approach proved to be more efficient in computational homogenization problems. 
Recently, \citet{faust2026empirical} developed a method, which uses a cluster-wise linearisation of the constitutive response around one cluster-wise reference point, at which stresses are evaluated.

All of these hyperreduction methods are tied to the full system used to generate the snapshots. 
Applying hyperreduction at the component level in substructured systems makes the approach spatially adaptive: component ROMs can be reused and assembled in different system configurations without recomputation. 
Because each substructure has fewer degrees of freedom than the full model, the offline cost is reduced. 
The design of the offline sampling strategy is challenging because the reduced models must handle varying boundary and interface coupling conditions.


To the authors' knowledge, the oldest works on component-wise MOR for nonlinear solid mechanics problems are \citet{barbicRealtimeLargedeformationSubstructuring2011, kimPhysicsbasedCharacterSkinning2012} in the field of computer graphics. 
\citet{barbicRealtimeLargedeformationSubstructuring2011} coupled the substructures using a penalty method. 
They solve the Newton-Euler equations for each substructure, where the rigid body motions are treated as independent degrees of freedom (DOF). 
To reduce the computational cost, they apply the hyperreduction method developed by \citet{an2008optimizing} to each substructure.
\citet{kimPhysicsbasedCharacterSkinning2012} treat the rotation using the gradient and the hessian of the rotation matrix. 
They couple the substructures for open-loop systems using Featherstone's algorithm \citep{featherstone1984robot}. 
Both methods from computer graphics do not fulfill the tied contact conditions exactly. 
\citet{zhou_proper_2018} use a penalty method for coupling and POD modes of the substructures. 
In structural dynamics, there exist some methods that extend the well-established Craig-Bampton method for geometric nonlinearities \citep{kuether_modal_2016,kuether_modal_2017,bui_reduced-order_2024}. 
\citet{kuether_modal_2016,kuether_modal_2017}, which extends the modal equations by using a quadratic and a cubic term. 
The nonlinear coefficients are computed by evaluating the nonlinear forces for the modal displacements. 
\citet{bui_reduced-order_2024} uses static modal derivatives to capture the nonlinearities and a Taylor series expansion of the nonlinear forces. 

Related contributions for nonlinear thermal problems were made by 
\citet{ebrahimiHyperreducedReducedBasis2024a,ebrahimiOnlineadaptiveHyperreducedReduced2026}, who propose a component-wise hyperreduction method. 
They treated interface and interior degrees of freedom (DOFs) separately and developed a component-wise empirical quadrature method based on the work by \citet{yano2019lp}. 
In the offline stage, they generated component snapshots from randomly sampled systems and tested the method on different systems assembled from the same components.

The present work extends the method developed by \citet{ritzert2025component}, where the substructures are reduced by a POD method and the coupling is done by a mortar method  that fulfills the tied contact conditions exactly. 
The DOFs of the substructures are split into slave-side interface DOFs, master-side interface DOFs and interior DOFs. 
Only the interior and master side interface DOFs are reduced, the slave side interface DOFs are projected onto the reduced space of the master side interface DOFs.
The mortar coupling has the advantage that non-matching interfaces can be handled (see e.g. \citep{wohlmuth_discretization_2001,puso_3d_2004,laursen_mortar_2012,popp_contact_2018}).
The novelty of this paper is that we extend the method by a component-wise ECSW hyperreduction method. 
The reduced sets of elements and the corresponding weights are computed in the offline stage for each substructure type individually and do not require any information about the coupling conditions.  
Additionally, we show that finite rotations can be represented by a linear combination of 9 modes, which can be derived from the general description of the rigid body rotation of a 3D body.  
The advantage of this approach is the simplicity compared to the treatment of rotations by \citet{kimPhysicsbasedCharacterSkinning2012} and \citet{barbicRealtimeLargedeformationSubstructuring2011}.

\paragraph{Research gap} 
Existing hyperreduction methods for nonlinear solid mechanics are mostly constructed for one fixed, fully assembled system. 
This limits their reuse in modular simulations, where the same component may appear in different assemblies, boundary conditions, or material models. 
A key open question is therefore whether hyperreduced component models can be trained independently and still be assembled reliably into new nonlinear systems.

\paragraph{Aim of this study} 
This study develops a component-wise hyperreduction framework for nonlinear solid mechanics. Component level POD bases and ECSW element weights are computed offline for individual substructures and then reused in different assemblies coupled by mortar mesh tying.
The aim is to show that such hyperreduced components can serve as transferable building blocks while retaining accuracy and computational efficiency under changing assemblies, boundary conditions, and material behavior.

\section{Component-wise model order reduction}\label{sec:equations} 

The component-wise model order reduction method (CWMOR) (see \citep{ritzert2025component}) is based on a mortar tied-contact full-order model where the slave-side contact DOFs and the Lagrange multipliers are condensed out.    
The remaining DOFs are projected into their respective substructure reduced spaces to obtain a reduced system of equations. 
In this chapter, we briefly explain the full-order model and the CWMOR method. 
We use a different notation than in \citet{ritzert2025component} to have a more compact notation for the further derivations.
\subsection{Full-order model}\label{sec:FOM} 


The full-order model (FOM) is governed by a tied-contact problem discretized using the finite element method.
\subsubsection*{Weak formulation}
The weak form of the problem is 
\begin{equation}\label{eq:saddlePoint} 
\begin{aligned}
    \sum_{i=1}^{n_s} \delta g^i_{\rm int}(\bu,\ddot \bu,\delta\bu)  + \sum_{j=1}^{n_c} \delta g^j_{c,u}(\Blambda,\delta\bu_c^1,\delta \bu_c^2) &= 0 \\
    \sum_{j=1}^{n_c} \delta g^j_{c,\lambda} ( \bu_c^1,\bu_c^2,\delta \Blambda) &= 0 
\end{aligned}
\end{equation} 
where $\delta g^i_{\rm int}$ is the virtual work done by the internal forces of substructure $i$ and $\delta g^j_{c,u}$ and $\delta g^j_{c,\lambda}$ are the virtual work contributions of the tied-contact conditions of interface $j$. 
The virtual work done by the internal forces of each substructure $i$ with its domain $\Omega^i_0$, the inertia forces, and the external forces is 
\begin{equation}\label{eq:internal_energy}
    \delta g_{\rm int}^i = \int_{\Omega^i_0} \left( \bS : \delta \bE\, +\rho_0 \ddot \bu \cdot \delta \bu  \, - \rho_0 \bb \cdot \delta \bu \right) \di V  - \int_{\Gamma_t} \delta \bu \cdot \bm{t} \, \di A,
\end{equation}
where $\bS$ is the second Piola-Kirchhoff stress tensor, $\delta \bE$ is the variation of the Green-Lagrange strain tensor, $\rho_0$ is the mass density in the reference configuration, $\bb$ is the body force per unit mass and $\bm{t}$ is the traction vector on the Neumann boundary $\Gamma_t$.
The virtual work contributions of the tied contact conditions at an interface $j$ are defined as
\begin{equation}
\label{eq:weakcontact0}
    \delta g^j_{c,u} =  \int_{\Gamma_c^{1,2}}  \left( \delta \bu_c^1 - \delta \bu_c^2 \right) \cdot \, \Blambda \, \di A 
\end{equation}
\begin{equation}
\label{eq:weakcontact1}
    \delta g^j_{c,\lambda} =  \int_{\Gamma_c^{1,2}}  \delta \Blambda \cdot \left( \bu_c^1 -  \bu_c^2 \right) \, \di A
\end{equation}
Here, $\Gamma_c^{1,2}$ is the contact interface between substructure 1 and 2, $\bu_c^1$ and $\bu_c^2$ are the displacements on the slave and master side of the interface, respectively, and $\Blambda$ are the Lagrange multipliers that enforce the tied contact condition.

\subsubsection*{Discretization}
We use the finite element method to discretize the problem in space and the Newmark-$\beta$ method for the time discretization, producing a nonlinear residual vector $\bG^i$ for each substructure. 
The residual vector is the sum of the internal forces, the inertial forces, and the external forces
\begin{equation}\label{eq:substructure_residual} 
    \bG^i =  \bM_{\rho}^i \ddot \bU^i + \bR^i(\bU^i)  - \bF_{\rm ext}^i
\end{equation}
where $\bM_{\rho}^i$ is the mass matrix, $\bR^i$ is the vector of internal forces, and $\bF_{\rm ext}^i$ is the vector of external dead forces.
The current nodal velocity and acceleration vectors $\dot \bU^i$ and $\ddot \bU^i$ are computed by the Newmark-$\beta$ method 
\begin{equation}\label{eq:Newmark} 
    \begin{aligned}
    \dot \bU^i &= \frac{\gamma}{\beta \Delta t} \left( \bU^i - \bU^i_{n} \right) - \frac{\gamma - \beta}{\beta} \dot \bU^i_{n} - \Delta t \frac{\gamma - 2\beta}{2\beta} \ddot \bU^i_{n} \\
    \ddot \bU^i &= \frac{1}{\beta \Delta t^2} \left( \bU^i - \bU^i_{n} \right) - \frac{1}{\beta \Delta t} \dot \bU^i_{n} - \frac{1 - 2\beta}{2\beta} \ddot \bU^i_{n}
    \end{aligned}
\end{equation}
where $\beta$ and $\gamma$ are the Newmark parameters, that we chose as $\beta = 0.25$ and $\gamma = 0.5$ in this article. 
 
Because of the nonlinear material behavior and the finite strain kinematics, the residual vector $\bG^i$ is a nonlinear function of the displacements $\bU^i$. 
The system is solved by the Newton-Raphson method, where the displacement vector is updated iteratively by $\bU^i \leftarrow \bU_n^i + \Delta \bU^i$ until convergence is reached. 
In each iteration, the following linear system has to be solved for the displacement increment $\Delta \bU^i$ and the Lagrange multipliers $\BLambda^j$
\begin{align}\label{eq:NewtonRaphson0}
    \sum_{i=1}^{n_s} \left( \bG_i(t,\bU_n^i) + \bK_i(\bU_n^i) \, \Delta \bU_i \right)+ \sum_{j=1}^{n_c} \left({\bD_j}^T \BLambda^j_n - {\bM_j}^T \BLambda^j_n \right) &= \bm 0 \\ 
    \sum_{j=1}^{n_c} \left(\bD_j \, \Delta\bU_{c,j}^1 - \bM_j \, \Delta\bU_{c,j}^2 \right) &= \bm 0 
    \label{eq:NewtonRaphson1}
\end{align} 
The matrices $\bD^j$ and $\bM^j$ stem from the finite element discretization of the tied contact conditions using the mortar method (see e.g.\cite{wohlmuth_discretization_2001, puso_3d_2004,laursen_mortar_2012,popp_contact_2018}). 
The mortar method allows for non-matching meshes on the contact interface.   
The tangential stiffness matrix $\bK^i$ is the Jacobian of the residual vector $\bG^i$ with respect to the displacements $\bU^i$.
For the dynamic case, the Newmark-beta time discretization results in the following expression for the tangential stiffness matrix
\begin{equation}\label{eq:substructure_stiffness} 
    \bK^i = \frac{\partial \bG^i}{\partial \bU^i} = \frac{1}{\beta \Delta t^2} \bM_{\rho}^i + \bK^i_R(\bU^i).  
\end{equation}
Here, $\bK^i_R$ is the partial derivative of the internal forces $\bR^i$ with respect to the displacements $\bU^i$.

In the following, we consider the case of two substructures with one interface, to have a simple notation for the further derivations. 
The master substructure will be denoted by a superscript $M$ and the slave substructure by a superscript $S$.
The extension to more substructures and interfaces is straightforward.
The linear system that has to be solved in each iteration of the Newton-Raphson method, is then given by

\begin{equation}\label{eq:firstLGS}
    \begin{bmatrix}
\bK_{II}^M  & \bK_{IC}^M & \bm{0} & \bm{0} & \bm{0} \\ 
\bK_{CI}^M & \bK_{CC}^M  & \bm{0} &  \bm{0} & - \bM^T \\ 
\bm{0} &\bm{0} & \bK_{II}^S  & \bK_{IC}^S & \bm{0} \\ 
\bm{0} &\bm{0} &  \bK_{CI}^S  & \bK_{CC}^S & \bD^T \\ 
\bm{0} & -\bM  & \bm{0}  & \bD & \bm{0}
\end{bmatrix}
\begin{bmatrix} 
\Delta \bU_I^M \\ \Delta \bU_C^M \\ \Delta \bU_I^S  \\ \Delta \bU_C^S \\ \Delta \BLambda
\end{bmatrix} = -
\begin{bmatrix} 
\bG_I^M  \\ \bG_C^M \\ \bG_I^S \\ \bG_C^S \\ \bm 0 
\end{bmatrix} 
\end{equation}

The DOFs of each substructure are separated into internal DOFs (denoted by the subscript $I$) and contact DOFs (denoted by the subscript $C$). 
The stiffness matrix and the residual vector are separated into blocks corresponding to the internal and contact DOFs. 
The mortar matrices $\bD$ and $\bM$ are only defined for the contact interface DOFs.

The system of equations \ref{eq:firstLGS} can be condensed by expressing the slave side contact DOFs by the master side contact DOFs
\begin{equation}\label{eq:interfaceCoupling}
    \Delta \bU_C^S = \bD^{-1}\bM \, \Delta \bU_C^M =  \bP \, \Delta \bU_C^M
\end{equation} 
Additionally, the Lagrange multiplier increments $\BLambda$ can be expressed as 
\begin{equation}\label{eq:disLagrangeMult}
    \Delta\BLambda = - \bD^{-T} \left( \bG_C^S + \bK_{CI}^S \, \Delta \bU_I^S  + \bK_{CC}^S \, \Delta \bU_C^S \right)
\end{equation} 
To evaluate \Cref{eq:interfaceCoupling} and \Cref{eq:disLagrangeMult}, we need the inverse of the mortar matrix $\bD$. 
By choosing dual shape functions for the Lagrange multipliers the mortar matrix $\bD$ becomes diagonal and is therefore easy to invert, as described e.g. in the works by \cite{wohlmuth_discretization_2001,popp_contact_2018,ritzert2025component}.   
With the above equations, the system of equations \Cref{eq:firstLGS} can be condensed to 

\begin{equation}\label{eq:condensedLGS}
    \underbrace{
\begin{bmatrix}
\bK_{II}^M  & \bK_{IC}^M & \bm{0} \\ 
\bK_{CI}^M  & \bK_{CC}^M + \bP^T \bK_{CC}^S \bP & \bP^T \bK_{CI}^S \\ 
\bm{0}  & \bK_{IC}^S \, \bP & \bK_{II}^S \\ 
\end{bmatrix}}_{\coloneqq \bK_{\rm cond}}
\underbrace{\begin{bmatrix} 
\Delta \bU_I^M  \\ \Delta \bU_C^M \\ \Delta \bU_I^S
\end{bmatrix}}_{\coloneqq \Delta \bU_{\rm cond}} = -
\underbrace{\begin{bmatrix} 
\bG_I^M \\ \bG_C^M + \bP^T \bG_C^S \\ \bG_I^S 
\end{bmatrix}}_{\coloneqq\bG_{\rm cond}}
\end{equation}
where the discrete interface coupling operator $\bP = \bD^{-1}\bM$ is used. 

For the application of MOR methods it is advantageous to rewrite the condensed system of \Cref{eq:condensedLGS} using a transformation matrix $\bT$.
The transformation matrix $\bT$ maps the condensed displacement vector $\Delta \bU_{\rm cond}$ to the original displacement vector $\Delta \bU$ by
\begin{equation}\label{eq:trafoU}
    \begin{bmatrix} \bU_I^M \\ \bU_C^M \\ \bU_I^S \\ \bU_C^S \end{bmatrix} = 
    \underbrace{\begin{bmatrix}
\bI & \bm{0} & \bm{0} \\ 
\bm{0} & \bI & \bm{0} \\
\bm{0} & \bm{0} & \bI \\ 
\bm{0} & \bP & \bm{0} 
\end{bmatrix}}_{\bT}
\begin{bmatrix} \bU_I^M \\ \bU_C^M \\ \bU_I^S \end{bmatrix}
\end{equation} 

The transpose of the transformation matrix $\bT^T$ maps the original residual vector $\bG$ to the condensed residual vector $\bG_{\rm cond}$ by 
\begin{equation}\label{eq:trafoG}
    \begin{bmatrix} \bG_I^M \\ \bG_C^M + \bP^T \bG_C^S \\ \bG_I^S  \end{bmatrix} = 
    \underbrace{\begin{bmatrix}
\bI & \bm{0} & \bm{0} & \bm{0} \\ 
\bm{0} & \bI & \bm{0} & \bP^T\\
\bm{0} & \bm{0} & \bI & \bm{0} \\ 
\end{bmatrix}}_{\bT^T}
\begin{bmatrix} \bG_I^M \\ \bG_C^M \\ \bG_I^S \\ \bG_C^S \end{bmatrix}
\end{equation} 
The residual vector $\bG$ contains the residuals of both substructures $S$ and $M$.
The condensed stiffness matrix $\bK_{\rm cond}$ can be obtained by 
\begin{equation}\label{eq:trafoK}
    \bK_{\rm cond} = 
    \begin{bmatrix}
        \bI & \bm{0} & \bm{0} & \bm{0} \\ 
        \bm{0} & \bI & \bm{0} & \bP^T\\
        \bm{0} & \bm{0} & \bI & \bm{0} \\ 
    \end{bmatrix}
    \begin{bmatrix} 
        \bK_{II}^M  & \bK_{IC}^M & \bm{0} & \bm{0}  \\ 
        \bK_{CI}^M & \bK_{CC}^M  & \bm{0} &  \bm{0}  \\ 
        \bm{0} &\bm{0} & \bK_{II}^S  & \bK_{IC}^S  \\ 
        \bm{0} &\bm{0} &  \bK_{CI}^S  & \bK_{CC}^S  
    \end{bmatrix} 
    \begin{bmatrix}
        \bI & \bm{0} & \bm{0} \\ 
        \bm{0} & \bI & \bm{0} \\
        \bm{0} & \bm{0} & \bI \\ 
        \bm{0} & \bP & \bm{0} 
    \end{bmatrix}
\end{equation}
The transformation matrix $\bT$ will later be used to derive the reduced system of equations.

\subsection{Component-wise model order reduction }\label{sec:ROM}  
In this article, we derive a component-wise hyperreduction method for nonlinear solid mechanics problems. 
This hyperreduction method is an extension of the component-wise model order reduction method that was introduced in our previous work\cite{ritzert2025component}. 
The idea of the component-wise model order reduction method is to compute a projection matrix for each substructure and to use these bases to derive a reduced system of equations for the condensed system (\ref{eq:condensedLGS}).

In this work, the projection matrices are computed for each substructure separately by using the POD method on a matrix composed  of displacement snapshots, see e.g. \citep{chatterjee2000introduction}.
For each substructure $i$, we compute $l$ snapshots of the displacements $\bU^i$ in an offline stage by solving the substructure problems for different loading and boundary conditions. 
These snapshots are collected in a snapshot matrix
$$
    \bS^i_{\rm Snap} = [\bU_1^i, \bU_2^i, \dots, \bU_l^i].
$$ 
To compute the projection matrix $\BPsi^i$ for substructure $i$, the snapshot matrix is decomposed by the singular value decomposition (SVD). 
\begin{equation}
    \bS^i_{\rm Snap} = \bm{V}^i \bm{X}^i \bm{W}^i
\end{equation}
The first $m_i$ left singular vectors are used as the projection matrix $\BPsi^i$ for substructure $i$ 
\begin{equation}
    \BPsi^i = \left[ \bV_1^i, \bV_2^i, \dots, \bV_{m_i}^i \right].
\end{equation}
With these substructure projection matrices $\BPsi^i$, we can derive a reduced system of equations for the full-order condensed system \Cref{eq:condensedLGS}.

Each projection matrix $\BPsi^i$ is split into internal and master contact DOF parts $\BPsi_I^i$ and $\BPsi_C^i$. 
All parts of the projection matrix that correspond to the slave-side contact DOFs are set to zero, since these DOFs are expressed by the master side contact DOFs.
To apply Dirichlet boundary conditions, the corresponding rows of the projection matrix are also set to zero. 
Considering the case of two substructures the condensed displacement vector is approximated as 
\begin{equation}\label{eq:approxU} 
    \begin{bmatrix} \bU_I^M \\ \bU_C^M \\ \bU_I^S \end{bmatrix} \approx 
    \begin{bmatrix}
        \BPsi_I^M & \bm{0} \\
        \BPsi_C^M  & \bm{0} \\
        \bm{0}  & \BPsi_I^S 
    \end{bmatrix}
    \begin{bmatrix} \ba^M \\ \ba^S \end{bmatrix} 
\end{equation}
Here, $\ba^M$ and $\ba^S$ are the reduced displacement vectors of the master and slave substructure, respectively.
Using the transformation matrix $\bT$ from \Cref{eq:trafoU} the original displacement vector can be approximated as
\begin{equation}\label{eq:approxUorig}
    \begin{bmatrix} \bU_I^M \\ \bU_C^M \\ \bU_I^S \\ \bU_C^S \end{bmatrix} \approx 
    \begin{bmatrix}
        \bI & \bm{0} & \bm{0} \\ 
        \bm{0} & \bI & \bm{0} \\
        \bm{0} & \bm{0} & \bI \\ 
        \bm{0} & \bP & \bm{0} 
    \end{bmatrix}
    \begin{bmatrix}
        \BPsi_I^M & \bm{0} \\
        \BPsi_C^M  & \bm{0} \\
        \bm{0}  & \BPsi_I^S 
    \end{bmatrix}
    \begin{bmatrix} \ba^M \\ \ba^S \end{bmatrix} 
    = 
    \underbrace{
    \begin{bmatrix}
        \BPsi_I^M & \bm{0} \\
        \BPsi_C^M  & \bm{0} \\
        \bm{0}  & \BPsi_I^S \\
        \bP \BPsi_C^M  & \bm{0}
    \end{bmatrix}}_{ \coloneqq \BPhi}
    \begin{bmatrix} \ba^M \\ \ba^S \end{bmatrix} 
    =  \BPhi \, \ba
\end{equation}
The global projection matrix $\BPhi$ maps the reduced displacement vector $\ba$ to the original displacement vector $\bU$. 
For systems with many substructures, it becomes a sparse matrix, since the projection matrices of each substructure only have nonzero entries corresponding to the DOFs of this substructure and the neighboring master substructures.

We can compute the reduced stiffness matrix $\hat \bK_{\rm cond} = \BPhi^T \bK \BPhi$ and the reduced residual vector $\hat \bG_{\rm cond} = \BPhi^T \bG$ by a Galerkin projection of the global stiffness matrix and residual with the global projection matrix $\BPhi$. 
In matrix notation, the reduced condensed stiffness matrix is given by 
\begin{equation}\label{eq:reducedK} 
    \hat \bK_{\rm cond} = 
    \begin{bmatrix}
        {\BPsi_I^M}^T & {\BPsi_C^M}^T & \bm{0} &  {\BPsi_C^M}^T \bP^T  \\
        \bm{0}  & \bm{0}  & {\BPsi_I^S}^T & \bm{0} 
    \end{bmatrix} 
    \begin{bmatrix} 
        \bK_{II}^M  & \bK_{IC}^M & \bm{0} & \bm{0}  \\ 
        \bK_{CI}^M & \bK_{CC}^M  & \bm{0} &  \bm{0}  \\ 
        \bm{0} &\bm{0} & \bK_{II}^S  & \bK_{IC}^S  \\ 
        \bm{0} &\bm{0} &  \bK_{CI}^S  & \bK_{CC}^S  
    \end{bmatrix} 
    \begin{bmatrix}
        \BPsi_I^M & \bm{0} \\
        \BPsi_C^M  & \bm{0} \\
        \bm{0}  & \BPsi_I^S \\
        \bP \BPsi_C^M  & \bm{0}
    \end{bmatrix}
\end{equation}
and the reduced condensed residual vector can be written as
\begin{equation}\label{eq:reducedR} 
    \hat \bG_{\rm cond} = 
    \begin{bmatrix}
        {\BPsi_I^M}^T & {\BPsi_C^M}^T & \bm{0} &  {\BPsi_C^M}^T \bP^T  \\
        \bm{0}  & \bm{0}  & {\BPsi_I^S}^T & \bm{0} 
    \end{bmatrix}  
    \begin{bmatrix} 
        \bG_I^M \\ \bG_C^M  \\ \bG_I^S  \\ \bG_C^S \end{bmatrix}
\end{equation}
Alternatively, the reduced quantitities can be written as a sum over all elements $e$ of the system, which is the form that is used for the implementation of the hyperreduction method in the next section, i.e.
\begin{equation}\label{eq:reducedGsum} 
    \hat \bG_{\rm cond} = \sum_{e \in \cE} \BPhi_e^T \bG_e
\end{equation} 
\begin{equation}\label{eq:reducedKsum} 
    \hat \bK_{\rm cond} = \sum_{e \in \cE} \BPhi_e^T \bK_e \BPhi_e
\end{equation}
Here, $\BPhi_e$, $\bG_e$, and $\bK_e$ denote the parts of the global projection matrix $\BPhi$, residual vector $\bG$, and stiffness matrix $\bK$, associated with element $e$.
The set of all elements $\cE$ is the union of the sets of elements of each substructure $\cE = \bigcup_i^{n_s} \cE^i$. 

The reduced residual and reduced stiffness matrix contain internal forces, inertia forces. 
Additionally, the reduced residual contains external forces. 
Only the internal forces $\bR$ and the tangential stiffness matrix $\bK$ depend nonlinearly on the displacements and have to be updated in each Newton-Raphson iteration. 
We split the reduced residual into three parts
\begin{equation}\label{eq:splitR} 
    \hat \bG_{\rm cond} = \sum_{e \in \cE} \BPhi_e^T \bR_e  +\sum_{e \in \cE} \BPhi_e^T \bM_{\rho, e} \BPhi_e \ddot \ba -  \sum_{e \in \cE} \BPhi_e^T \bF_{\rm ext,e} = \hat \bR_{\rm cond} + \hat \bM_{\rho, \rm cond} \ddot \ba - \hat \bF_{\rm ext,cond} 
\end{equation}
The reduced internal force part $\hat \bR_{\rm cond}$ is updated in each iteration of the Newton-Raphson method, while the reduced inertia part $\hat \bM_{\rho, \rm cond}$ can be precomputed in the offline stage or at the beginning of the online stage.
It is multiplied with the reduced acceleration vector $\ddot \ba$ to obtain the inertia forces in the reduced system.
The reduced acceleration vector $\ddot \ba$ is computed by projecting the Newmark-$\beta$ equations \Cref{eq:Newmark} into the reduced space by the global projection matrix $\BPhi$.
Analogously, the reduced stiffness matrix is also split into a stiffness part, related to the internal forces, and an inertia part, related to the reduced mass matrix
\begin{equation}\label{eq:splitK} 
    \hat \bK_{\rm cond} = \sum_{e \in \cE} \BPhi_e^T \bK_{R,e} \BPhi_e + \frac{1}{\beta \Delta t^2 } \sum_{e \in \cE} \BPhi_e^T \bM_{\rho, e} \BPhi_e  = \hat \bK_{R,\rm cond} +  \frac{1}{\beta \Delta t^2 } \hat \bM_{\rho, \rm cond}.
\end{equation}
This split into internal forces and inertia forces is used for the implementation of the hyperreduction method, since only the part related to the internal forces has to be updated in each iteration of the Newton-Raphson method.

\section{Component-wise hyperreduction by energy conserving sampling and weighting}\label{sec:ECSW}

The reduced system derived in section 2.2 still depends on the original size of the problem because all elements have to be evaluated to assemble the reduced internal force vector $\hat \bR_{\rm cond}$ and the reduced tangential stiffness matrix $\hat \bK_{\rm cond}$. 
Hyperreduction techniques such as DEIM \citep{chaturantabut_nonlinear_2010}, ECSW \citep{an2008optimizing, farhat_structure-preserving_2015}, or ECM \citep{hernandez_dimensional_2017}  reduce this effort by evaluating only a reduced number of elements to  approximate $\hat \bR_{\rm cond}$ and $\hat \bK_{\rm cond}$. 
In this work we use ECSW for each component because in contrast to DEIM it does not require additional snapshots for the internal force vectors.
DEIM does not preserve the structure of the equation system and can lead to numerical instabilities, especially for large rotations because of the split of the forces into linear and nonlinear parts \cite{rutzmoser_model_2018}. 
Compared to ECM, ECSW was easier to implement into our existing finite element code, since it works with element level quantities, while ECM works with Gauss point level quantities. 
In preliminary investigations, ECM does not have significant advantages over ECSW in terms of accuracy and computational efficiency. 
Since ECSW has advantages in terms of implementation since it works with element level quantities and is widely used in the literature,  ECSW is selected as the hyperreduction method for this work.

The component-wise ECSW approximation is not identical to applying ECSW to the assembled system, because the online projection operator contains interface contributions induced by the mortar mesh coupling. 
The approximation is expected to be accurate if the offline boundary parametrization spans the interface deformation states encountered online.  
This assumption is tested in the numerical examples (\Cref{sec:examples}) by applying the same component weights to assemblies and loading conditions not used during training. 
In \Cref{sec:ECSW} we discuss essential ingredients of the standard ECSW method (\Cref{sec:ECSW}) and then present the component-wise extension in \Cref{sec:ECSWmodular}.

\subsection{Energy conserving sampling and weighting}\label{sec:ECSW} 
In our implementation we follow the descriptions of \citet{farhat_structure-preserving_2015} and \citet{rutzmoser_model_2018}. 
In this section we give a brief overview of the method; for more details the reader is kindly referred to the original articles. 
The idea of the ECSW method is to approximate the virtual work of the internal forces in the reduced system by a weighted sum of the element virtual work contributions of a reduced set of elements $\tilde \cE$.    
The total virtual work of the reduced system is the sum of the element virtual work contributions over the whole set of elements $\cE$. 
This approximation is written as
\begin{equation}\label{eq:energyECSW} 
    \delta \ba^T \hat \bR = \delta\ba^T  \sum_{e \in \cE} \BPsi_e^T   \bR_e \approx \delta\ba^T \sum_{e \in \tilde\cE} w_e \,   \BPsi_e^T \bR_e,
\end{equation}
where $w_e$ are the element weights and $\BPsi_e$ is the part of the projection matrix $\BPsi$ associated with element $e$. 
The virtual work equality \Cref{eq:energyECSW} has to hold for arbitrary virtual reduced displacements $\delta \ba$, leading to 
\begin{equation}\label{eq:forceECSW} 
    \sum_{e \in \cE} \BPsi_e^T  \bR_e \approx \sum_{e \in \tilde\cE} w_e \,  \BPsi_e^T \bR_e.
\end{equation}  
The weights $w_e$ can be computed by enforcing \Cref{eq:forceECSW} for a set of displacement snapshots \\ $\bm{S}_{\rm Snap} = \left[ \bm{U}_1, \bm{U}_2  , \dots, \bm{U}_l \right] $. 
In this work, we only use converged snapshots. 
Since we search for a sparse solution, we minimize the zero-norm of the weight vector $\lVert  \bw \rVert_0$, which leads to the constrained optimization problem
\begin{equation} 
    \bw^*=\textrm{arg min}   \: \lVert  \bw \rVert_0, \:\: \textrm{subject to} \: \lVert \bY \bw - \bb \rVert^2 \textrm{  and} \: \bw \ge 0. 
\end{equation} 
An exact solution of this is computationally infeasable, instead we use a greedy algorithm to compute an approximate solution. 
This algorithm, the sparse non-negative least squares algorithm (sNNLS), is described in the work by \cite{farhat_structure-preserving_2015}.
The algorithmsuccessively adds promising candidate elements, recomputes weights, and terminates when the convergence criterion 
\begin{equation}\label{eq:ECSWconvergence} 
    \lVert \bY \bw - \bb \rVert \le \tau \lVert \bb \rVert
\end{equation}
is reached.
The accuracy of the solution can be controlled by the tolerance $\tau$. 
A smaller value of $\tau$ leads to a larger number of nonzero entries in $\bw$. Consequently, a higher number of elements has to be evaluated in the online stage. 

The quantities $\bY,\,\bw$ and $\bb$ are  
\begin{align}
    \bY = 
    &\begin{bmatrix} 
    \BPsi_1^T \bR_1 ( \BOmega \bU_1) & \dots & \BPsi_{n_{\cE}}^T \bR_{n_{\cE}} (\BOmega \bU_1) \\ 
    \vdots & \ddots & \vdots \\ 
    \BPsi_1^T \bR_1 (\BOmega \bU_l) & \dots & \BPsi_{n_{\cE}}^T \bR_{n_{\cE}} (\BOmega \bU_l) \\ 
    \end{bmatrix}\\ 
    \bw = 
    &\begin{bmatrix} 
    w_1 \\ \vdots \\ w_{n_{\cE}}
    \end{bmatrix} \\ 
    \bb = 
    &\begin{bmatrix} 
    \sum_{e\in \cE} \BPsi_e^T \bR_e (\BOmega \bU_1) \\ \vdots \\ 
    \sum_{e\in \cE} \BPsi_e^T \bR_e (\BOmega \bU_l)
    \end{bmatrix}
\end{align}
The operator $\BOmega$ is defined as $\BOmega = \BPsi  \BPsi^T $. It computes the orthogonal projection of a snapshot $\bU$ onto the subspace spanned by $\BPsi$. 
In each row of the matrix $\bY$, the element contributions $\BPsi_e^T \bR_e$ of all elements $e$ are evaluated for one snapshot. 
In the vector $\bb$ the sum of all element contributions $\sum_{e\in \cE} \BPsi_e^T \bR_e$ is evaluated for each snapshot. 
The weight vector $\bw$ is computed by the sNNLS such that the weighted sum of the element contributions in $\bY$ approximates the sum of all element contributions in $\bb$ for all snapshots.  
Each row in the equation system $\bY \bw = \bb$ is \Cref{eq:forceECSW} evaluated for one snapshot.

With the weights $w$, the hyperreduced internal force vector $\tilde \bR$ and the hyperreduced tangential stiffness matrix $\tilde \bK_R$ can be constructed
\begin{align} 
    \tilde \bR &\approx \sum_{e \in \tilde\cE} w_e \BPsi_e^T \bR_e \\ 
    \tilde \bK_R &\approx \sum_{e \in \tilde \cE} w_e \BPsi_e^T \bK_e \BPsi_e
\end{align}
The above equations are weighted sums over the reduced set of elements $\tilde \cE $, which contains all elements with nonzero weights: $\tilde \cE = \{e \in \cE \vert w_e > 0 \}$. 
In the following, hyperreduced quantities that are approximated by ECSW are denoted by a tilde $\tilde\bullet$.

\subsection{Application to the component-wise MOR method}\label{sec:ECSWmodular} 
In the standard ECSW method, the projection matrix $\BPsi$, the weights $w$, and the reduced set of elements $\bar \cE$ are computed for the whole system. 
In the component-wise model order reduction method, this cannot be done, since the tied-contact conditions and the Dirichlet conditions applied to the assembled system are not known beforehand. 
Additionally, there exist no snapshots for the whole system, which makes offline weight computation for the whole system impossible. 
We propose in this paper to compute the projection matrix, the weights, and the reduced sets of elements for each substructure separately from the substructure snapshots.  
The weights and the reduced sets of elements can be combined to get the weights and the reduced set of elements for the whole system. 
The latter are then used for the rapid online evaluation of the reduced internal force vector and the reduced tangential stiffness matrix.

For each substructure $i$, we compute a projection matrix $\BPsi^i$, the ECSW weights $\bw^i$ and the reduced set of elements $\bar \cE^i$ from the substructure snapshots. 
The reduced sets of elements of the substructures are then combined to get the reduced set of elements for the whole system $\tilde \cE = \bigcup_i^{n_s} \bar \cE^i$ and the weights are combined to get the weights for the whole system  $\bw = \begin{bmatrix} {\bw^1} & {\bw^2} & \hdots & {\bw^{n_s}} \end{bmatrix}$. 
With these weights, the reduced internal force vector, defined in \Cref{eq:reducedRtilde} and the reduced tangential stiffness matrix, defined in \Cref{eq:reducedKtilde} can be approximated by a weighted sum over the reduced set of elements $\tilde \cE$ as described in the previous section
\begin{equation}\label{eq:reducedRtilde} 
    \tilde \bR_{\rm cond} \approx \sum_{e \in \tilde\cE} w_e \BPhi_e^T \bR_e
\end{equation}
\begin{equation}\label{eq:reducedKtilde} 
    \tilde \bK_{R, \rm cond} \approx \sum_{e \in \tilde\cE} w_e \BPhi_e^T \bK_e \BPhi_e
\end{equation}
This is an approximation of the reduced internal forces and the reduced tangential stiffness parts derived in \Cref{eq:splitR} and \Cref{eq:splitK} by a weighted sum over the combined reduced set of elements $\tilde \cE$. 
The projection matrix $\BPhi$ is the global projection matrix defined in \Cref{eq:approxUorig} and contains the projection matrices of all substructures as well as the tied contact conditions. 
As a consequence, the slave side contact DOFs of all elements in the reduced set of elements are projected by the projection matrix of the master side.


\section{Computation of snapshots}\label{sec:snapshots} 

\subsection{Snapshot generation}\label{sec:snapshotMethod}
In this work, we compute the snapshots of the substructures by performing simulations of single substructures with parametrized Dirichlet boundary conditions at the possible interfaces and boundary faces. 
The snapshots are computed by a Latin Hypercube Sampling (LHS) (cf.\citep{mckay2000comparison,stein1987large} )  of the parameter space of the boundary conditions. 
This approach was also used by \citet{ritzert2025component}.

In other works, the snapshots are computed by simulating a small system of a few substructures with parametrized boundary conditions. 
This is sometimes called oversampling strategy, see e.g. \citep{diercks2023multiscale, ebrahimiHyperreducedReducedBasis2024a}.
The advantage of the oversampling strategy is that the snapshots are computed in a realistic setting.
The interface deformations are influenced by the neighboring substructures, which is also the case in the systems used in the online phase. 
The training systems have to be chosen such that different coupling scenarios are covered.
The disadvantage is that the snapshot generation is more expensive compared to computing snapshots of single substructures.
Computing snapshots of single substructures is computationally cheaper than the oversampling strategy, but the boundary conditions have to be parametrized to span the linear subspace in which the possible coupling scenarios lie.

In an assembled system, interface deformations are dominated by translations and rotations of the interfaces. Additionally, the interfaces undergo deformations in the normal- and tangential directions.
In this work, we compute the displacement boundary conditions by applying parametrized translations and rotations to the possible interfaces of the substructure.
To allow for normal and tangential deformations, we use two different strategies. 
The first strategy is applied for two-dimensional problems, it is used later in \Cref{sec:samplingKreuz}. There we use three different sets of boundary conditions for each sample. 
The first set suppresses both normal deformation and tangential deformations. 
The second set permits normal deformation, while the third set permits tangential deformations.
For three-dimensional problems we use the second strategy (see \Cref{sec:samplingRing}), where elastic blocks are attached to the interfaces of the substructure. 
The displacement boundary conditions are applied to the outer edges of these blocks instead of directly to the interfaces.
As a result, the interfaces can deform in the normal direction and contract in tangential directions.
For three-dimensional problems we have two tangential and one normal boundary condition direction that can all be turned on and off. 
The first strategy leads to seven different sets of boundary conditions for each sample, because all possible combinations of activated and inactivated boundary conditions are used.
This would significantly increase the computational cost of the snapshot generation.

A challenge is to determine the range of the parameters for the boundary conditions. The ranges can be chosen by testing a small assembled system, that contains different coupling scenarios.
In nonlinear problems, not just the shape of the deformation but also the magnitude of the deformation has to be represented by the snapshots. 
There is no clear rule to choose the parameter ranges for the boundary conditions, but some rules of thumb can be given.
In many systems that experience large deformations, the biggest part of the displacement is caused by rigid body motions of the substructures. 
The substructure deformations are often dominated by tension while shearing is small.
The reason for this is that at free surfaces, no shear reactions are possible. 
This leads to a moment reaction that rotates the substructure towards tensile configurations. 
If the substructures are slender or have free surfaces the parametrization should represent this effect. 
Tangential displacements and rotations of the interfaces of the substructures should be smaller than the tensile normal displacements.
Choosing a larger parameter space leads to a higher number of substructure POD modes, that often do not represent the deformations that occur in the online phase.

\subsection{Rigid body modes}\label{sec:RBM} 
When multiple substructures are connected the deformation of one substructure can lead to rigid body translations and rotations of another substructure. 
The substructure modes need to include modes that can represent arbitrary rigid body motions. 
For small strains the rigid body motions can be represented by the six modes of the null space of the stiffness matrix. 
These modes are the three translation modes and the three linearized rotation modes.
The linearized rotation modes contain displacements specific to a single rotation angle. 
If other rotations are approximated by this mode they would change the volume and shape of the body. 
This additional deformation has to be represented by additional modes. 

\begin{figure}[h!]
    \centering
\begin{tikzpicture}

\definecolor{crimson2143940}{RGB}{214,39,40}
\definecolor{darkgray176}{RGB}{176,176,176}
\definecolor{darkorange25512714}{RGB}{255,127,14}
\definecolor{steelblue31119180}{RGB}{31,119,180}
\definecolor{forestgreen4416044}{RGB}{44,160,44}

\begin{groupplot}[group style={group size=3 by 1}, width = 6.2 cm  ]
\nextgroupplot[
legend cell align={left},
legend style={
  fill opacity=0.8,
  draw opacity=1,
  text opacity=1,
  at={(0.5,-0.15)},
  anchor=north,
  draw=none
},
x grid style={darkgray176},
xmin=-1.8, xmax=1.8,
xtick style={color=black},
y grid style={darkgray176},
ymin=-1.8, ymax=1.8,
ytick style={color=black},
axis equal image
]
\addplot [semithick,dashed , steelblue31119180, mark=*, mark size=2, mark options={solid}]
table {%
-1 -1
1 -1
1 1
-1 1
-1 -1
};
\addlegendentry{reference square}
\addplot [semithick, steelblue31119180, mark=*, mark size=2, mark options={solid}]
table {%
0.366025403784438 -1.36602540378444
1.36602540378444 0.366025403784438
-0.366025403784438 1.36602540378444
-1.36602540378444 -0.366025403784438
0.366025403784438 -1.36602540378444
};
\addlegendentry{rotated square}

\nextgroupplot[
legend cell align={left},
legend style={
  fill opacity=0.8,
  draw opacity=1,
  text opacity=1,
  at={(0.5,-0.15)},
  anchor=north,
  draw=none
},
x grid style={darkgray176},
xmin=-1.8, xmax=1.8,
xtick style={color=black},
y grid style={darkgray176},
ymin=-1.8, ymax=1.8,
ytick style={color=black},
axis equal image
]
\addplot [semithick,dashed, steelblue31119180, mark=*, mark size=2, mark options={solid}, forget plot]
table {%
-1 -1
1 -1
1 1
-1 1
-1 -1
};
\addplot [semithick, steelblue31119180, mark=*, mark size=2, mark options={solid}, forget plot]
table {%
0.366025403784438 -1.36602540378444
1.36602540378444 0.366025403784438
-0.366025403784438 1.36602540378444
-1.36602540378444 -0.366025403784438
0.366025403784438 -1.36602540378444
};
\addplot [semithick, darkorange25512714, mark=*, mark size=2, mark options={solid}]
table {%
0.239230484541326 -1.61961524227066
1.61961524227066 0.239230484541326
-0.239230484541326 1.61961524227066
-1.61961524227066 -0.239230484541326
0.239230484541326 -1.61961524227066
};
\addlegendentry{fit by one rotation mode}

\nextgroupplot[
legend cell align={left},
legend style={
  fill opacity=0.8,
  draw opacity=1,
  text opacity=1,
  at={(0.5,-0.15)},
  anchor=north,
  draw=none
},
x grid style={darkgray176},
xmin=-1.8, xmax=1.8,
xtick style={color=black},
y grid style={darkgray176},
ymin=-1.8, ymax=1.8,
ytick style={color=black},
axis equal image
]
\addplot [semithick,dashed, steelblue31119180, mark=*, mark size=2, mark options={solid}, forget plot]
table {%
-1 -1
1 -1
1 1
-1 1
-1 -1
};
\addplot [semithick, steelblue31119180, mark=*, mark size=2, mark options={solid}, forget plot]
table {%
0.366025403784438 -1.36602540378444
1.36602540378444 0.366025403784438
-0.366025403784438 1.36602540378444
-1.36602540378444 -0.366025403784438
0.366025403784438 -1.36602540378444
};
\addplot [semithick, darkorange25512714, mark=*, mark size=2, mark options={solid}]
table {%
0.366025403784438 -1.36602540378444
1.36602540378444 0.366025403784438
-0.366025403784438 1.36602540378444
-1.36602540378444 -0.366025403784438
0.366025403784438 -1.36602540378444
};
\addlegendentry{fit by two modes}
\addplot [semithick,dotted, crimson2143940, mark=*, mark size=2, mark options={solid}]
table {%
0.154700538379251 -1.57735026918963
1.57735026918963 0.154700538379251
-0.154700538379251 1.57735026918963
-1.57735026918963 -0.154700538379251
0.154700538379251 -1.57735026918963
};
\addlegendentry{rotation part}
\addplot [semithick, dotted, forestgreen4416044, mark=*, mark size=2, mark options={solid}]
table {%
-0.788675134594813 -0.788675134594813
0.788675134594813 -0.788675134594813
0.788675134594813 0.788675134594813
-0.788675134594813 0.788675134594813
-0.788675134594813 -0.788675134594813
};
\addlegendentry{volumetric part}
\end{groupplot}

\end{tikzpicture}
    \caption{Two dimensional example of a rotation and the representation by modes. The left plot shows the initial and the rotated configuration. The center plot shows the fit by a single rotation mode, which leads to an increase in area. The right plot shows the fit by two modes, where the second mode is a volumetric expansion/contraction mode that counteracts the increase in area.} 
    \label{fig:rotations}
\end{figure}
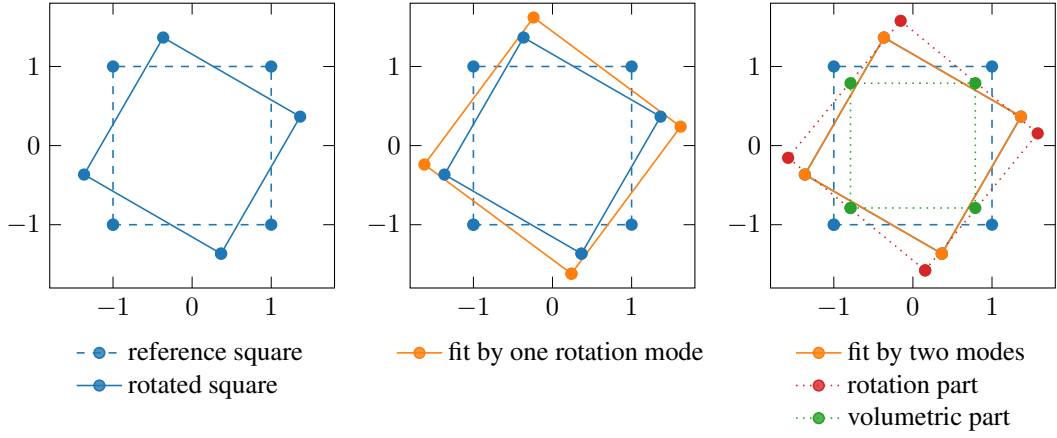

Let us consider a 2D example to illustrate this. 
In \Cref{fig:rotations} a square with edge length 2 is rotated around the center by an angle of $\alpha = 60^\circ$. 
In the left plot, one can see the square in its initial and rotated configuration. 
The plot in the center shows the fit of the rotation by a single linearized rotation mode around the angle $\bar\alpha = 36.87^\circ$.     
It can be seen that the rotation cannot be fitted and that the volume of the rectangle increases. 
The right plot shows the fit by two modes, where we used a volumetric expansion as the second mode.
With these two modes we can represent the rotation with machine precision. 
The two red squares are the contributions of the rotation mode and the volumetric expansion mode. 
The rotation mode leads to an increase in area which is counteracted by the shrinkage of the volumetric mode.  
In 2D all rigid body deformations can be represented by two translation modes, one rotation mode and one volumetric expansion mode. 

The same principle holds also for general rotations in 3D, the rotation modes introduce deformations of the body when used for different angles and rotation axes. 
These additional deformations have to be counteracted by a linear combination of deformation modes. 
For general rotations these deformation modes are not as easy to interpret as in the 2D case. 
They still contain volumetric expansion modes, but also other deformations like shearing and stretching. 

The nine modes that can describe rotation can be derived from the general description of the rigid body rotation of a 3D body.
The displacement resulting from a rigid body rotation can be written as
\begin{equation}\label{eq:general_rotation}
    \bu(\bp,\bX) = \bR(\bp)\bX
\end{equation}
where $\bX$ is the position vector of a point in the body and $\bR$ is a matrix that maps the position vector to the displacement vector $\bu$. 
The vector $\bp$ contains the parameters for a general rotation, e.g.\ the rotation angle and the rotation axis or Euler angles. 
It is related to a rotation matrix $\bQ(\bp)$ by $\bR(\bp) = \bQ(\bp) - \bm{1}$.
For general rotations the matrix $\bR(\bp)$ has 9 independent entries, because  each entry depends nonlinearly on the parameters in $\bp$. 
Therefore, we can define nine spatial modes that can express any rotational deformation. 

These modes can be computed by expressing the displacement resulting from rotations using nine spatial modes 
\begin{equation}\label{eq:rotation_modes} 
    \bu(\bp,\bX) = \sum_{i=1}^9 \bV_i(\bX) a_i(\bp) 
\end{equation}
where $\bV_i$ are the rotation modes and $a_i$ are the amplitudes of the modes depending on the rotation parameters $\bp$. 
The modes $\bV_i$ depend only on the position $\bX$ and not on the rotation parameters $\bp$. 
We choose the modes $\bV_i$ to be linearly dependent on the position $\bX$: $ \bV_i(\bX) = \bA_i \bX$, where $\bA_i$ are constant matrices. 
By setting \Cref{eq:general_rotation} and \Cref{eq:rotation_modes} equal to each other and by inserting the above definitions for $\bV_i$, we can derive 
\begin{equation} 
    \left( \bR(\bp) - \sum_{i=1}^9 \bA_i a_i(\bp) \right) \bX = 0.
\end{equation} 
The constant matrices $\bA_i$ are the constant modal coefficients of an affine decomposition of $\bR(\bp)$, with $a_i(\bp)$ being the corresponding nonlinear coefficients.
When we write the rotation matrix using index notation, we obtain 
\begin{equation} 
    \sum_{j=1}^3 \sum_{k=1}^3 R_{jk}(\bp) \be_j \otimes \be_k = \sum_{i=1}^9 \bA_i a_i(\bp)
\end{equation}
with $\be_j$ being the unit vector in $j$-th direction. 
From that equation we can directly read off the constant matrices $\bA_i = \be_j \otimes \be_k$ and the amplitudes $a_i(\bp) = R_{jk}(\bp)$, with $i = k+3(j-1)$. 
For each node of the mesh, the modes required to express an arbitrary rotation can be written as 
\begin{equation} 
    \bV_i(\bX) = \left( \be_j \otimes \be_k\right)  \bX = X_k \be_j, \qquad i = k+3(j-1).
\end{equation}

For the 2D case, the same principle holds, but the rotation matrix has only two independent entries, which leads to two rotation modes. 
In 2D the rotation is described by the single rotation angle $\alpha$. 
The displacements resulting from a rotation are
\begin{equation} 
    \bu(\alpha,\bX) = \bR(\alpha)\bX = \begin{bmatrix} \cos(\alpha) - 1 & -\sin(\alpha) \\ \sin(\alpha) & \cos(\alpha) - 1 \end{bmatrix} \begin{bmatrix} X\\ Y \end{bmatrix}.
\end{equation}
The same displacement can be represented as 
\begin{equation} 
    \bu(\alpha,\bX) = \begin{bmatrix} 1 & 0 \\ 0 & 1 \end{bmatrix} \bX (\cos(\alpha) - 1) + \begin{bmatrix} 0 & -1 \\ 1 & 0 \end{bmatrix} \bX \sin(\alpha) 
\end{equation}
From which we can read off the two rotation modes 
\begin{equation} 
\bV_1(\bX) = \begin{bmatrix} 1 & 0 \\ 0 & 1 \end{bmatrix}  \bX = \begin{bmatrix}  X\\ Y\end{bmatrix} \quad \text{and} \quad \bV_2(\bX) = \begin{bmatrix} 0 & -1 \\ 1 & 0 \end{bmatrix} \bX = \begin{bmatrix} -Y \\ X \end{bmatrix}.
\end{equation}

Additionally to the rotation modes, we need three translation modes in 3D and two translation modes in 2D to represent all rigid body motions. 
These can simply be defined as constant modes $\bV_i = \be_i$ with $\be_i$ being the unit vector in $i$-th direction.

In matrix notation, the full 3D rigid body projection matrix can be defined for each node $\bX^i$ as  
\begin{equation} 
    \BPsi_{\textrm{RBM}}^i(\bX) = 
    \begin{bmatrix}  
        1 & 0 & 0    & X^i & 0 & 0     & Y^i & 0 & 0    & Z^i & 0   & 0 \\
        0 & 1 & 0    & 0 & X^i & 0     & 0 & Y^i & 0    & 0   & Z^i & 0 \\
        0 & 0 & 1    & 0 & 0 & X^i    & 0 & 0 & Y^i     & 0   & 0   & Z^i  
    \end{bmatrix} 
\end{equation}

In 2D, the nodal rigid body projection matrix is
\begin{equation} 
    \BPsi_{\textrm{RBM}}^i(\bX) = 
    \begin{bmatrix}  
        1 & 0    & X^i & -Y^i \\
        0 & 1    & Y^i & X^i 
    \end{bmatrix}
\end{equation}

The full projection matrix $\BPsi_{\textrm{RBM}} = \left[ (\BPsi_{\textrm{RBM}}^1)^T \, (\BPsi_{\textrm{RBM}}^2)^T, \dots, \, (\BPsi_{\textrm{RBM}}^k)^T  \right]^T$ is obtained by stacking the nodal projection matrices $\BPsi_{\textrm{RBM}}^i$ for all $k$ nodes and normalizing the columns. 
These projection matrix are not unique: any basis which spans the same linear subspace could be used instead. 
ONe possible choice might be a POD basis computed from snapshots of pure rigid body motions, which results in exactly 9 non-vanishing singular values in 3D and 4 non-vanishing singular values in 2D.
For the illustrations in \Cref{fig:rotations},  we used a different projection matrix.

\subsection{Snapshot centering}\label{sec:snapsPreprocessing} 
As we will show in the numerical examples, it can be an advantage to remove the rigid body mode part of the snapshots before computing the POD basis.
The substructure snapshots often contain rigid body motions. 
Each snapshot $\bU_i$ can be expressed as the sum of the deformation part and the rigid body contribution $ \bU_i =\bU_i^{\textrm{Defo}} + \bU_i^{\textrm{RBM}}  $.
In order to get pure deformation snapshots without rigid body motions the rigid body motion contributions need to be subtracted
\begin{equation} 
    \bU_i^{\textrm{Defo}} = \bU_i - \bU_i^{\textrm{RBM}}.
\end{equation}
The rigid body part can be expressed with the rigid body motion POD basis shown in \Cref{sec:RBM}
\begin{equation} 
    \bU_i^{\textrm{RBM}} \approx \BPsi_{\textrm{RBM}}\,  \ba_i^{\textrm{RBM}},
\end{equation}
where $\ba_i^{\textrm{RBM}}$ is the amplitude vector of the rigid body motion modes. 
The amplitudes can be computed by a linear least squares fit 
    $\min_{\ba_i^{\textrm{RBM}} \in \mathbb{R}^{9}} \| \BPsi_{\textrm{RBM}} \ba_i^{\textrm{RBM}} - \bU_i \|^2. $
The solution of this minimization is given by the normal equation 
    $\BPsi_{\textrm{RBM}}^T \BPsi_{\textrm{RBM}} \, \ba_i^{\textrm{RBM}} = \BPsi_{\textrm{RBM}}^T \bU_i.$ 
Since $\BPsi_{\textrm{RBM}}$ is an orthonormal basis, the product $\BPsi_{\textrm{RBM}}^T \BPsi_{\textrm{RBM}} = \bm{1} \in \mathbb{R}^{9}$ gives the identity. 
The rigid body motion displacement vector $\bU_i^{\textrm{RBM}}$ can directly be computed 
\begin{equation} 
    \bU_i^{\textrm{RBM}} = \BPsi_{\textrm{RBM}}\,\BPsi_{\textrm{RBM}}^T \bU_i 
\end{equation}
This gives the deformation part of the snapshot 
\begin{equation}\label{eq:centering} 
    \bU_i^{\textrm{Defo}} = \bU_i - \BPsi_{\textrm{RBM}}\,\BPsi_{\textrm{RBM}}^T \bU_i.
\end{equation}
The POD basis is than computed from a modified snapshot matrix 
\begin{equation}\label{eq:DefoSnaps} 
    \bS_{\rm Defo} = \begin{bmatrix} \bU_1^{\textrm{Defo}} & \bU_2^{\textrm{Defo}} & \dots & \bU_l^{\textrm{Defo}} & \bS_{\textrm{RBM}}  \end{bmatrix},
\end{equation}
that contains all deformation contributions of the snapshot and the pure rigid body motion snapshots $\bS_{\textrm{RBM}}$.

Alternatively, the rigid body part of each snapshot could be computed by a Kabsch algorithm \citep{kabsch1976solution}, which computes the optimal rotation and translation to align two point clouds. 
But since we use modes to describe the finite rigid body motions in the online phase, we use the same modes to compute the rigid body part of the snapshots. 
It should be noted that $\bU_i^{\textrm{RBM}}$ contains not only the rigid body motion but also deformations. 


\section{Numerical examples}\label{sec:examples}

In this article, we test different aspects of the component-wise hyperreduction method using two main numerical examples. 
First, we investigate the effect of the snapshot centering explained in \Cref{sec:snapsPreprocessing} and the accuracy of the component-wise ECSW method (\Cref{sec:ECSWmodular}) for 2D and 3D systems. 
Then we demonstrate the performance of the method using larger systems with many substructures, where we show its predictive capabilities for quasi-static and dynamic problems. 
In \Cref{sec:dynExample}, we show that we can use quasi-static elastic snapshots to predict the behavior of a nonlinear hyper-viscoelastic dynamic system.
In all numerical simulations two different types of substructures are used.
In the first subsection we explain the boundary conditions and the parameter space for the snapshot sampling of both substructures.

\subsection{Snapshot sampling}\label{sec:snapshot_sampling}
In this section, we use two different substructures, which were already used in our previous publication \citet{ritzert2025component}. 
In the following, the boundary conditions and parameter spaces for the sampling process of both substructures are explained. 
For all snapshots a compressible Neo-Hookean material model with the Lame constants  $\lambda = 14907 \:\rm MPa, \mu = 34783 \: \rm MPa$ is used. 
The Helmholtz free energy of the Neo-Hookean material model is 
\begin{equation}
    \psi(\bC) = \frac{\mu}{2} \left( \tr(\bC) -3 - \ln(\det(\bC)) \right) + \frac{\lambda}{4} \left( \det(\bC) -1 - \ln(\det(\bC)) \right)
\end{equation}

\subsubsection{Substructure A}\label{sec:samplingKreuz} 

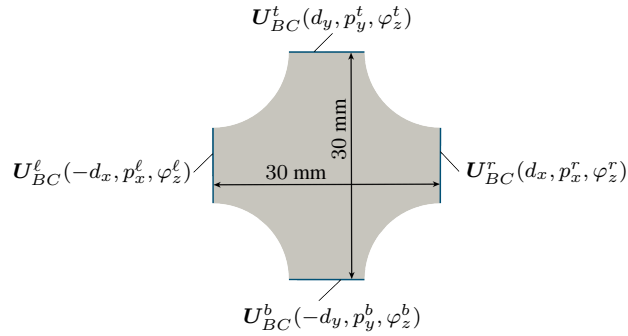
\begin{figure}[hptb]
    \centering
    {\footnotesize
    \begin{tikzpicture}
        \node[inner sep=0pt] (pic) at (0,0) {\includegraphics[width=0.25\textwidth]{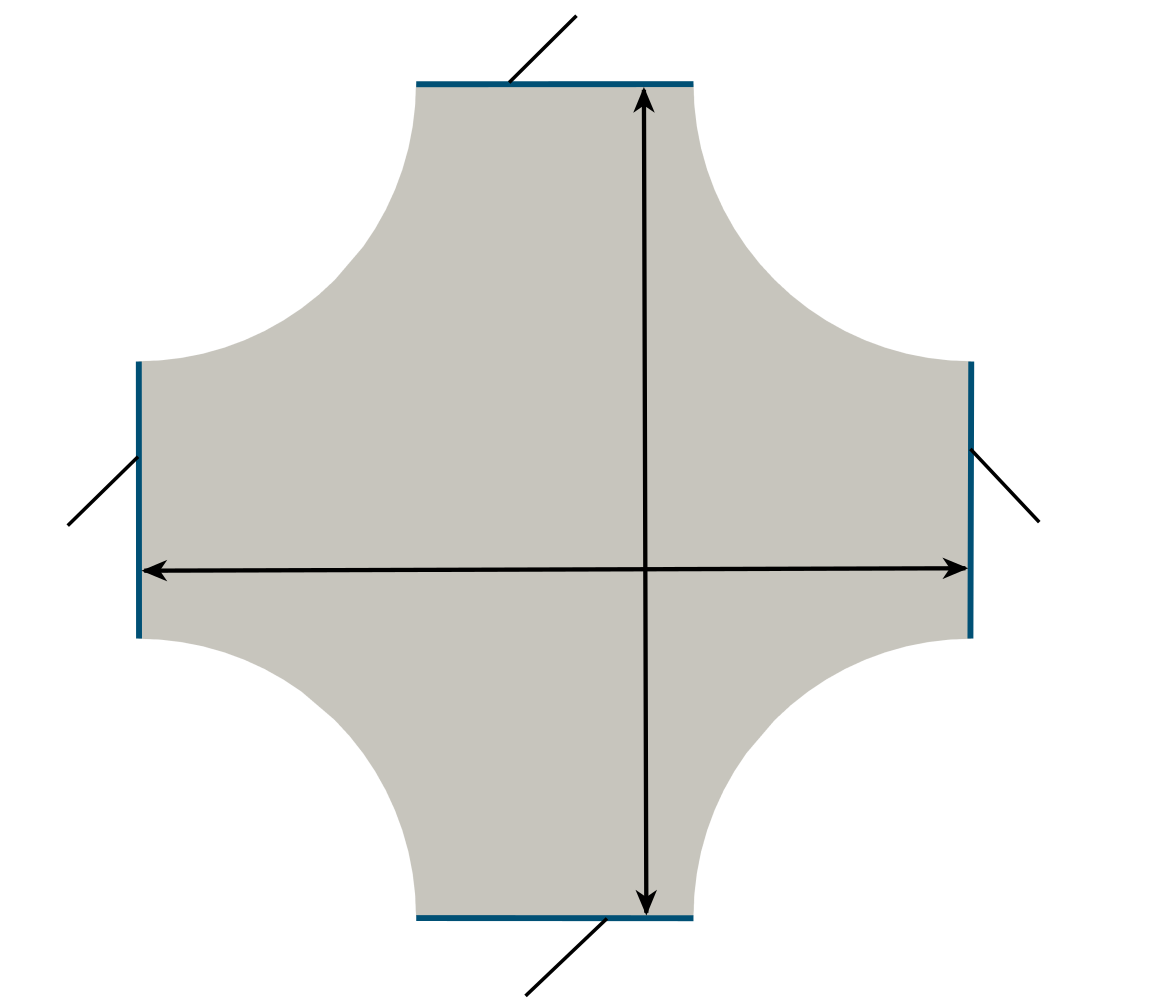}}; 
 
    \node[inner sep=0pt] (a) at ($(pic.south)+(-0.8*3.8,0.8*2.15)$) 
                {$\bU_{BC}^\ell (-d_{x}, p_{x}^\ell,  \phi_{z}^\ell)$};

    \node[inner sep=0pt] (a) at ($(pic.south)+(0.8*3.55,0.8*2.15)$) 
                {$\bU_{BC}^r (d_{x}, p_{x}^r, \phi_{z}^r)$};

    \node[inner sep=0pt] (a) at ($(pic.south)+(0,0.75*4.99)$) 
                {$\bU_{BC}^t (d_{y}, p_y^t, \phi_{z}^t)$};

    \node[inner sep=0pt] (a) at ($(pic.south)+(0,-0.2)$) 
                {$\bU_{BC}^b (-d_{y},p_{y}^b, \phi_{z}^b)$};
    
    \node[inner sep=0pt] (a) at ($(pic.center)+(-0.6*0.8,-0.6*0.15)$) 
                {$30 $ mm};

    \node[inner sep=0pt, rotate = 90] (a) at ($(pic.center)+(0.6*0.15,0.6*0.8)$) 
                {$30 $ mm};

    \end{tikzpicture}
    }
    \caption{Illustration of the snapshot parametrization. Figure adapted from \citet{ritzert2025component}.}
    \label{fig:parametrisierung}
\end{figure}

The first substructure is shown in \Cref{fig:parametrisierung}.
The edges marked in blue are the possible interfaces for boundary conditions, and tied contact interfaces.  
The same substructure with the same snapshot sampling procedure was already used in our work \citet{ritzert2025component}. 
We will briefly summarize the parametrization and boundary conditions of the sampling procedure. 

In the snapshot sampling, we apply Dirichlet boundary conditions on those edges. 
These are parametrized by three parameters on each edge:
The parameter $d_x$ or $d_y$ describes the displacement perpendicular to the edge, the parameter  $p_y^l$, $p_y^r$, $p_x^t$ or $p_y^b$ is responsible for the displacement parallel to the edge, and the parameter  $\phi_z^l,\,\phi_z^r,\, \phi_z^t,\,\phi_z^b $ parametrizes $x$- and $y$-displacements coming from a rotation of the edge.  
The parameters used in the sampling differ slightly from the ones used in \citet{ritzert2025component}. 
Here, we use
\begin{equation*} 
    \begin{aligned} 
        &d_x \in (-1.2 \,\mathrm{mm}  ,4 \,\mathrm{mm}  ) \qquad &&d_y \in (-1.2 \,\mathrm{mm},4 \,\mathrm{mm})  \\
        &p_x^l \in (-4 \,\mathrm{mm},4 \,\mathrm{mm}) \qquad &&\phi_z^l \in (-15^\circ,15^\circ)   \\ 
        &p_x^r \in (-4 \,\mathrm{mm},4 \,\mathrm{mm})  \qquad &&\phi_z^r \in (-15^\circ,15^\circ) \\ 
        &p_y^t \in (-4 \,\mathrm{mm},4 \,\mathrm{mm}) \qquad &&\phi_z^t \in (-15^\circ,15^\circ)  \\
        &p_y^b \in (-4 \,\mathrm{mm},4 \,\mathrm{mm}) \qquad &&\phi_z^b \in (-15^\circ,15^\circ)  
    \end{aligned}
\end{equation*} 


As described in \Cref{sec:snapshotMethod}, we consider three different boundary conditions cases. 
In the first case, we apply Dirichlet boundary conditions in tangential and normal direction of each edge.    
In the second and third case, only the tangential DOFs or the normal DOFs are prescribed.
This strategy allows for non-planar deformations of the interfaces. 
For a more detailed explanation of this choice the reader is referred to \citet{ritzert2025component}.

We generat 50 samples using LGS in the 10-dimensional parameter space.
Two load steps per sample and the three boundary condition cases yield 300 snapshots in total.

\subsubsection{Substructure B}\label{sec:samplingRing}


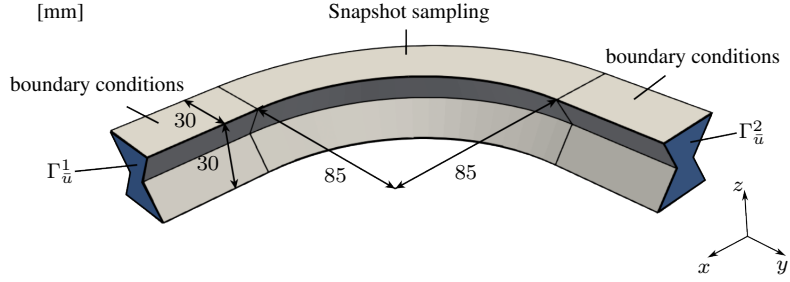
\begin{figure}[hptb]
    \centering
    {\footnotesize
    \begin{tikzpicture}
    \node[inner sep=0pt] (pic) at (0,0) {\includegraphics[width=0.6\textwidth]{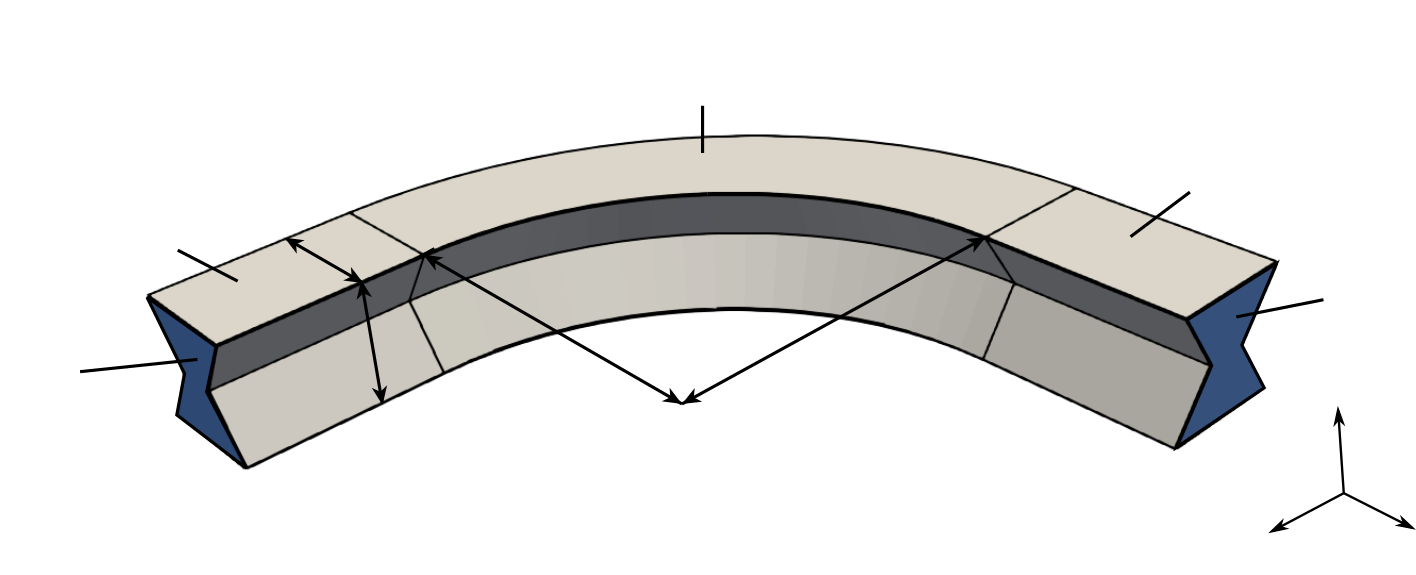} };
    \node[inner sep=0pt] at ($(pic.south) + (0.75*5.2,0.75*0.2)$) {$x$};
    \node[inner sep=0pt] at ($(pic.south) + (0.75*6.58,0.75*0.21)$) {$y$};
    \node[inner sep=0pt] at ($(pic.south) + (0.75*5.8,0.75*1.65)$) {$z$};
    \node[inner sep=0pt] at ($(pic.east) + (0.75*-0.6,0.75*-0.1)$) {$\Gamma_{\bar u}^2$};
    \node[inner sep=0pt] at ($(pic.west) + (0.75*0.5,0.75*-0.8)$) {$\Gamma_{\bar u}^1$};
    \node[inner sep=0pt] at ($(pic.north) + (0.75*0.0,0.75*-0.7)$) {Snapshot sampling};
    \node[inner sep=0pt] at ($(pic.north) + (0.75*-5.5,0.75*-2.0)$) {boundary conditions};
    \node[inner sep=0pt] at ($(pic.north) + (0.75*5.0,0.75*-1.5)$) {boundary conditions};
    \node[inner sep=0pt] at ($(pic.south) + (0.75*1,0.75*1.9)$) {$85$}; 
    \node[inner sep=0pt] at ($(pic.south) + (0.75*-1.3,0.75*1.85)$) {$85$}; 
    \node[inner sep=0pt] at ($(pic.west) + (0.75*3.1,0.75*-0.65)$) {$30$};
    \node[inner sep=0pt] at ($(pic.west) + (0.75*2.7,0.75*0.11)$) {$30$};
    \node[inner sep=0pt] at ($(pic.west) + (0.75*0.5,0.75*2.)$) {[mm]};
    \end{tikzpicture}
    }
    \caption{Boundary value problem for the snapshot computation, consisting of three substructures. The snapshots are collected for the central substructure. Figure adapted from \cite{ritzert2025component}}. 
    \label{fig:snapshot_ringstueck}
\end{figure}  

In \Cref{fig:snapshot_ringstueck}, we show the dimensions of the substructure as well as the boundary value problem used for the snapshot sampling. 
In this example, a parametric displacement \(\bU_{BC} = \bU_{BC}(d_x, \, d_y, \, d_z,  \,\phi_x, \, \phi_y, \, \phi_z ) \) is prescribed on both surfaces $\Gamma_{\bar u}^1$ and $\Gamma_{\bar u}^2$, where \( d_x, \, d_y, \, d_z \) define translational components along the respective coordinate axes, and \( \phi_x, \, \phi_y, \, \phi_z \) describe rotations around them. 
For boundary conditions on the surface $\Gamma_{\bar u}^1$, the parameter ranges used in the LHS are
\begin{equation*} 
    \begin{aligned} 
        &d_x \in (-36 \, \rm{mm},84\, \rm{mm} )  \qquad &&\phi_x \in (-70^\circ,70^\circ)   \\
        &d_y \in (-44 \, \rm{mm},36 \, \rm{mm})  \qquad &&\phi_y \in (-70^\circ,70^\circ)   \\
        &d_z \in (-150 \, \rm{mm},150 \, \rm{mm}) \qquad &&\phi_z \in (-40^\circ,40^\circ)  \\
    \end{aligned}
\end{equation*}


For each sampled parameter set, two simulations are performed: one with the displacement applied to $\Gamma_{\bar u }^1$ while $\Gamma_{\bar u }^2$ remains fixed, and another with the roles reversed. 
When the boundary conditions are applied on $\Gamma_{\bar u}^2$, the parameters are $d_x$ and $d_y$ are swapped, while the other parameters remain unchanged.

\subsection{Comparison between snapshots with and without snapshot centering}

\subsubsection{3D example}\label{sec:schlange} 

\begin{figure}[ht]
    \centering
    \def\svgwidth{0.7\textwidth}
    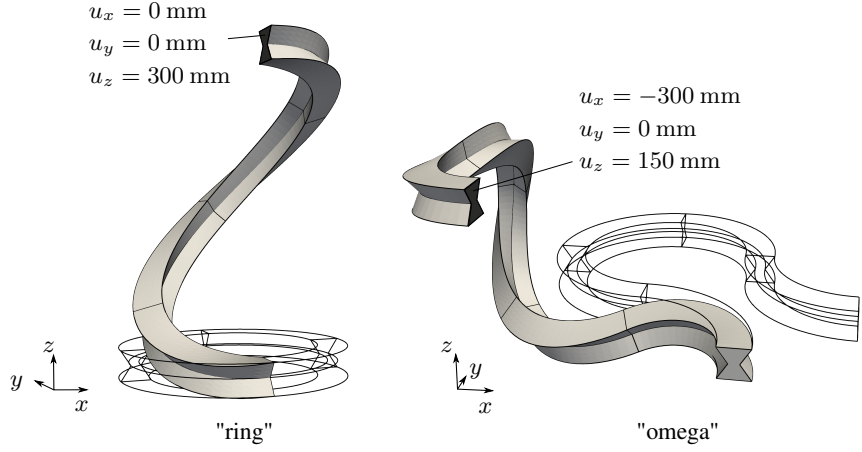
    \caption{Illustration of the geometry, and the boundary value problems for the 3D examples. Figure adapted from \cite{ritzert2025component} }.
    \label{fig:geometrySchlange}
\end{figure}

In this numerical example, we use substructure B in two different systems.
The boundary value problems are illustrated in \Cref{fig:geometrySchlange}. 
On the left, four components are assembled into a ring structure with a cut, with 15876 DOFs in total.  
The other system consists of five substructures with 19845 DOFs, forming an "Omega"-like shape. 
The displacement boundary conditions can be seen in \Cref{fig:geometrySchlange}.
In the following, we use the term "ring"-structure for the example on the left and the term "omega" for the structure shown on the right.   

\begin{figure}[h!] 
    \centering
    \input{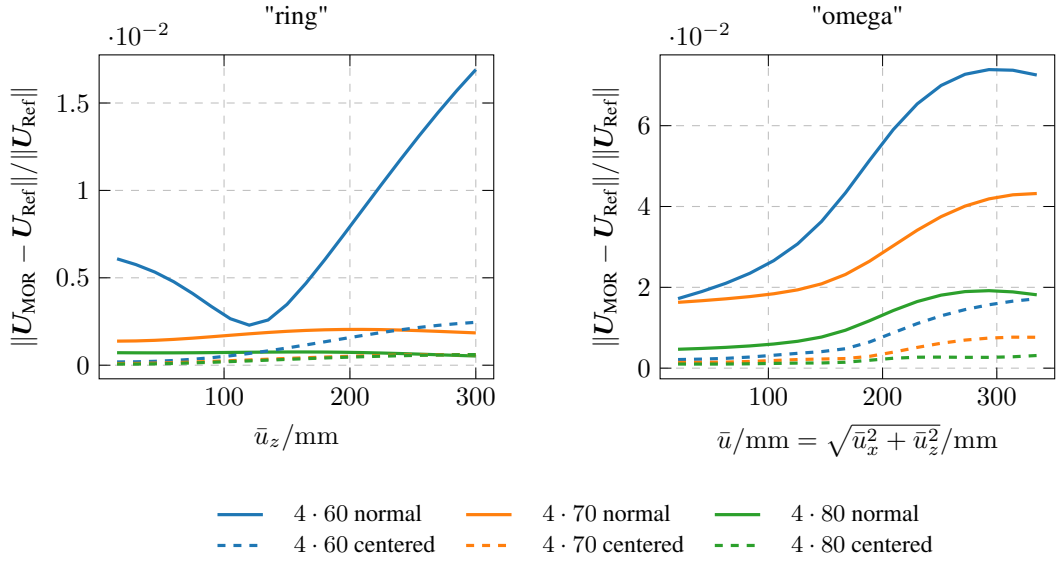}
    \caption{This plot shows the influence of the centering of the snapshots (see \Cref{sec:snapsPreprocessing}) on the accuracy of the solution for the "ring" system on the left and the "omega" system on the right. The relative displacement error is plotted over the applied displacement for different numbers of modes per substructure. } 
    \label{fig:ringOmega_centered_normal}
\end{figure}

In \Cref{fig:ringOmega_centered_normal} it can be seem, that the accuracy of the solution is improved when the snapshots are centered according to \Cref{sec:snapsPreprocessing}. 
In the centering the rigid body motion part computed from the rigid body modes are subtracted from each snapshot. 
The modes are computed from the pure deformation snapshots and the rigid body motion snapshots. 
For the "ring" system, the difference between the two methods becomes small for a higher number of modes per substructure. 
For 60 modes per substructure centering reduces the maximum error from $ 1.7\, \% $ to $ 0.24 \,\%$. 
For 80 modes per substructure the error is only reduced from $ 0.07\,\%$ to $ 0.06\,\%$. 
For the "omega" system, the difference is larger. 
For 80 modes per substructure the error is reduced from $1.9 \,\%$ to $0.3\,\%$. 
It can be concluded that the centering of the snapshots can increase the accuracy of the solution for small numbers of modes per substructure. 
If the number of modes is large enough to produce accurate results, the centering has only a small influence on the accuracy of the solution.

\begin{figure}[htbp]
    \centering
    \begin{tikzpicture}

  \definecolor{crimson2143940}{RGB}{214,39,40}
  \definecolor{darkgray176}{RGB}{176,176,176}
  \definecolor{darkorange25512714}{RGB}{255,127,14}
  \definecolor{forestgreen4416044}{RGB}{44,160,44}
  \definecolor{lightgray204}{RGB}{204,204,204}
  \definecolor{steelblue31119180}{RGB}{31,119,180}
  
  \begin{groupplot}[group style = {group name = group, group size = 2 by 1, horizontal sep = 2.2 cm}, width = 6.8 cm    ]  
      \nextgroupplot[
          legend cell align={left},
          legend style={font=\scriptsize },
          x dir=reverse,
          xlabel={$\tau $},
          xmajorgrids,
          xtick = {0.001, 0.01, 0.1}, 
          xticklabels={$10^{-3}$,$10^{-2}$,$10^{-1}$ },
          xmin=0.00079, xmax=0.125,
          xtick style={color=black},
          xmode=log,
          y grid style={darkgray176},
          ylabel={Error $ e $},
          ymajorgrids,
          ymin=0.00103004870818062, ymax=0.0313121596867666,
          ytick style={color=black},
          grid style = dashed
          ]
          \addplot [very thick, steelblue31119180, mark=*, mark size=2, mark options={solid}]
          table {%
          0.1 0.0228825168589674
          0.075 0.0294851640141498
          0.05 0.0246768448182424
          0.025 0.00600065410752342
          0.01 0.00606705284559362
          0.001 0.00436238367030727
          };
          \addlegendentry{$5\cdot70$}
          \addplot [very thick,dashed, steelblue31119180, forget plot]
          table {%
          0.1 0.00802720098224519
          0.075 0.00802720098224519
          0.05 0.00802720098224519
          0.025 0.00802720098224519
          0.01 0.00802720098224519
          0.001 0.00802720098224519
          };
          \addplot [very thick, darkorange25512714, mark=*, mark size=2, mark options={solid}]
          table {%
          0.1 0.0259492751772794
          0.075 0.0299357000968309
          0.05 0.0167199706660246
          0.025 0.00457754875417623
          0.01 0.00397636627205468
          0.001 0.00240650829811635
          };
          \addlegendentry{$5\cdot80$}
          \addplot [very thick,dashed, darkorange25512714, forget plot]
          table {%
          0.1 0.00390834535845294
          0.075 0.00390834535845294
          0.05 0.00390834535845294
          0.025 0.00390834535845294
          0.01 0.00390834535845294
          0.001 0.00390834535845294
          };

      \nextgroupplot[
          legend cell align={left},
          legend style={font=\scriptsize,
            at={(0.03,0.97)},
            anchor=north west
          },
          xlabel={$\tau $},
          x dir=reverse,
          xmajorgrids,
          xtick = {0.001, 0.01, 0.1}, 
          xticklabels={$10^{-3}$,$10^{-2}$,$10^{-1}$ },
          xmin=0.00079, xmax=0.125,
          xtick style={color=black},
          xmode=log,
          y grid style={darkgray176},
          ylabel={element ratio},
          ymajorgrids,
          ymin=-0.044, ymax=1.028,
          ytick style={color=black},
          ytick = {0, 0.25, 0.5,0.75, 1}, 
          grid style = dashed
          ]
          \addplot [very thick, steelblue31119180, mark=*, mark size=2, mark options={solid}]
          table {%
          0.1 0.0691666666666667
          0.075 0.07625
          0.05 0.0902083333333333
          0.025 0.110416666666667
          0.01 0.149166666666667
          0.001 0.311041666666667
          };
          \addlegendentry{$5\cdot70$}
          \addplot [very thick, darkorange25512714, mark=*, mark size=2, mark options={solid}]
          table {%
          0.1 0.074375
          0.075 0.084375
          0.05 0.0954166666666667
          0.025 0.124375
          0.01 0.166458333333333
          0.001 0.333333333333333
          };
          \addlegendentry{$5\cdot80$}
          
  \end{groupplot}

\end{tikzpicture}
    \caption{Influence of the tolerance $\tau$ on the accuracy of the solution of the "omega" system. On the left, the mean displacement error (\Cref{eq:mean_error}) is plotted over $\tau$. The dashed lines are the mean displacement error without ECSW. 
    On the right, the increase of the ratio of evaluated elements is plotted.} 
    \label{fig:omega_uerrECSW}
\end{figure}
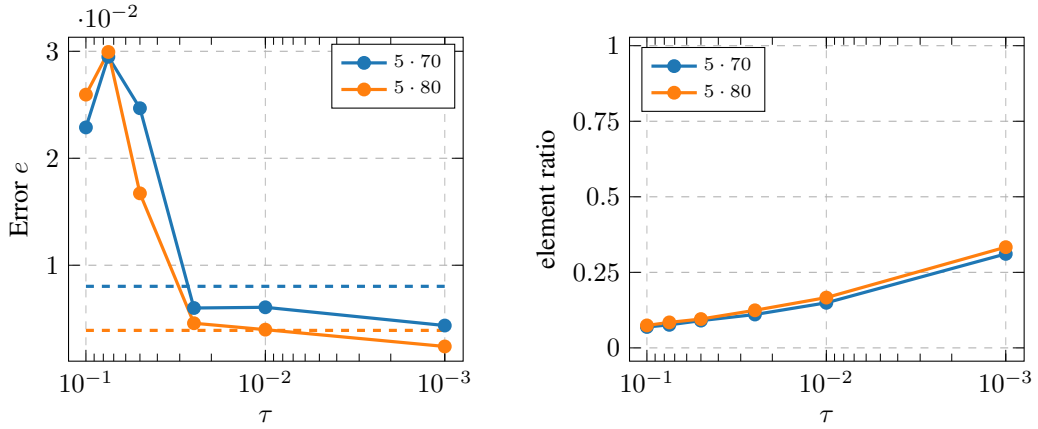

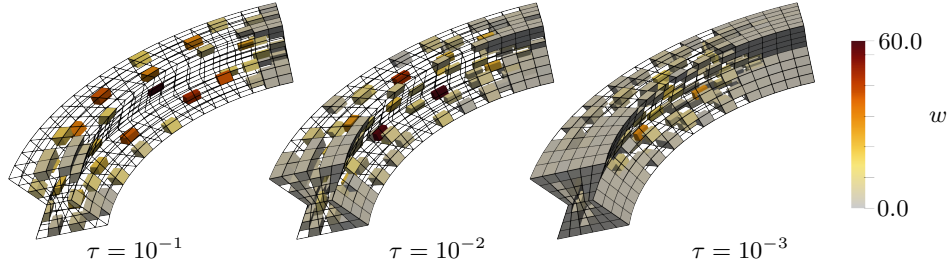
\begin{figure}[hptb]
    \centering
    {\small
    \begin{tikzpicture}
    \node[inner sep=0pt] (pic) at (0,0) {\includegraphics[width=0.7\textwidth]{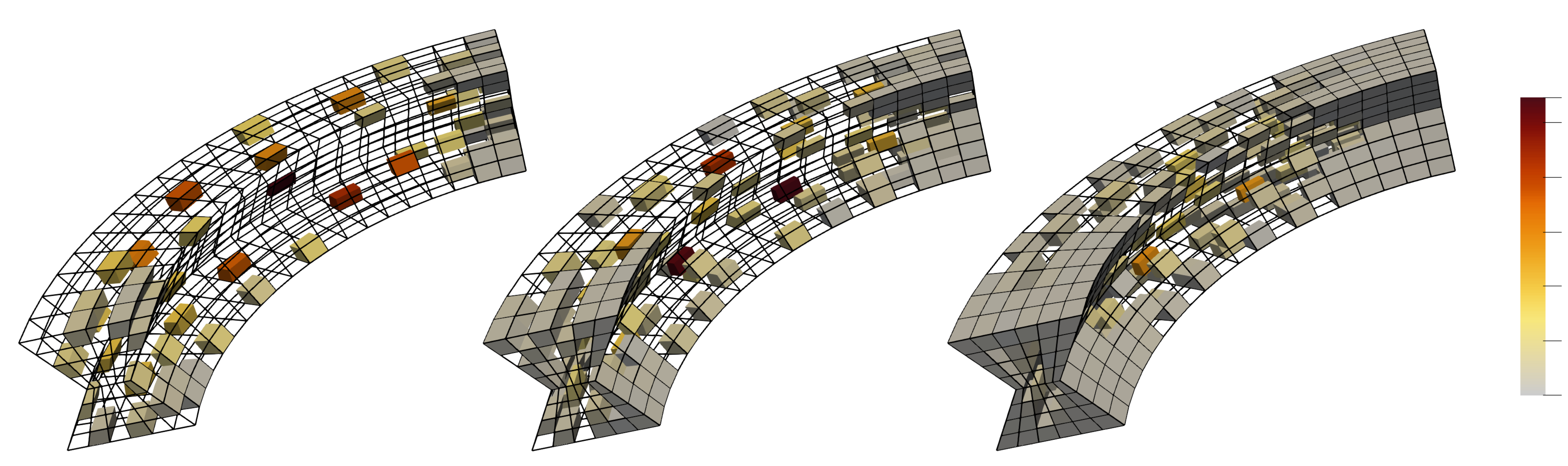} };
    \node[inner sep=0pt] at ($(pic.south) + (-4,0)$) {$ \tau= 10^{-1} $};
    \node[inner sep=0pt] at ($(pic.south) + (0,0)$) {$ \tau= 10^{-2} $};
    \node[inner sep=0pt] at ($(pic.south) + (4,0)$) {$ \tau= 10^{-3} $};
    \node[inner sep=0pt] at ($(pic.east) + (0.3,1.0)$) {$ 60.0 $};
    \node[inner sep=0pt] at ($(pic.east) + (0.2,-1.2)$) {$ 0.0 $};
    \node[inner sep=0pt] at ($(pic.east) + (0.8,0)$) {$ w $};
    \end{tikzpicture}
    }
    \caption{Visualization of the ECSW weights for the substructure used in the "ring" and "omega" systems. The colorbar shows the weights of the elements. The three subfigures show the weights for different tolerances $\tau$ computed with the 70 POD modes.  } 
    \label{fig:ring_ecsw_elements}
\end{figure}

In \Cref{fig:ring_uerrECSW,fig:omega_uerrECSW} we show the mean displacement error and the element ratio of evaluated elements to the total number of elements over the tolerance $\tau$ for both systems ("ring" and "omega"). 
The tolerance $\tau$ is the parameter controlling the convergence criterion (\Cref{eq:ECSWconvergence}) of the weight computation of the ECSW method.
The mean error is computed as the mean of the relative displacement error over all load steps 
\begin{equation}\label{eq:mean_error}
    e = \frac{1}{N_{\rm load}} \sum_{i=1}^{N_{\rm load}} \frac{\| \bU_{\rm MOR}^i - \bU_{\rm Ref}^i \|_2}{\| \bU_{\rm Ref}^i \|_2}. 
\end{equation}
Here, $N_{\rm load}$ is the number of load steps and $\bU_{\rm MOR}$ is either the POD solution or the ECSW solution.
For both systems the errors converge to the POD reduced solution. 
It can be seen, that smaller numbers of POD modes per substructure require smaller values of the tolerance $\tau$ to reconstruct the POD solution. 
For the "ring" example, $\tau=0.025$ for 60 modes per substructure gives accurate results with a difference to the POD mean error of $1.7 \cdot 10^{-3}$. 
For this hyperreduced computation $9.5 \%$ of all elements have to be evaluated.    
The difference of the ECSW solution for the "omega" structure is $2.3\cdot 10^{-3}$, where 70 modes per substructure and a tolerance of $\tau=0.025$ were used. 
This simulation required the evaluation of $11.0\%$ of all elements. 
The selected elements with their corresponding weights are visualized in \Cref{fig:ring_ecsw_elements} for three different values of $\tau$. 
The weights where computed with 70 POD modes per substructure.
It can be seen, that the selected elements are located predominantly at the interfaces of the substructure. 
In the regions with many ECSW elements, the weights are small compared to the highest weights. 
The elements at the boundary are neccessary, to ensure that the mesh tying conditions can be enforced.

\begin{figure}[htbp]
    \centering
    \begin{tikzpicture}

  \definecolor{crimson2143940}{RGB}{214,39,40}
  \definecolor{darkgray176}{RGB}{176,176,176}
  \definecolor{darkorange25512714}{RGB}{255,127,14}
  \definecolor{forestgreen4416044}{RGB}{44,160,44}
  \definecolor{lightgray204}{RGB}{204,204,204}
  \definecolor{steelblue31119180}{RGB}{31,119,180}
  
  \begin{groupplot}[group style = {group name = group, group size = 2 by 1, horizontal sep = 2.2 cm}, width = 6.8 cm    ]  
      \nextgroupplot[
          legend cell align={left},
          legend style={font=\scriptsize },
          x dir=reverse,
          xlabel={$\tau$},
          xmajorgrids,
          xtick = {0.001, 0.01, 0.1}, 
          xticklabels={$10^{-3}$,$10^{-2}$,$10^{-1}$ },
          xmin=0.00079, xmax=0.125,
          xtick style={color=black},
          xmode=log,
          y grid style={darkgray176},
          ylabel={Error $ e $},
          ymajorgrids,
          ymin=-0.000490320333229522, ymax=0.0167528454846601,
          ytick style={color=black},
          grid style = dashed
          ]
          \addplot [very thick, steelblue31119180, mark=*, mark size=2, mark options={solid}]
          table {%
          0.1 0.00613967099591864
          0.075 0.0159690652202106
          0.05 0.00967647940208367
          0.025 0.0029746269948749
          0.01 0.00138644751815476
          0.001 0.00121328787540176
          };
          \addlegendentry{$4\cdot60$}
          \addplot [very thick,dashed, steelblue31119180, forget plot]
          table {%
          0.1 0.00126379425669336
          0.075 0.00126379425669336
          0.05 0.00126379425669336
          0.025 0.00126379425669336
          0.01 0.00126379425669336
          0.001 0.00126379425669336
          };
          \addplot [very thick, darkorange25512714, mark=*, mark size=2, mark options={solid}]
          table {%
          0.1 0.00954153005666851
          0.075 0.00442712655080872
          0.05 0.00411147638885527
          0.025 0.00222389630930145
          0.01 0.00150049814990634
          0.001 0.000293459931220007
          };
          \addlegendentry{$4\cdot70$}
          \addplot [very thick,dashed, darkorange25512714, forget plot]
          table {%
          0.1 0.000400099251496335
          0.075 0.000400099251496335
          0.05 0.000400099251496335
          0.025 0.000400099251496335
          0.01 0.000400099251496335
          0.001 0.000400099251496335
          };
          \addplot [very thick, forestgreen4416044, mark=*, mark size=2, mark options={solid}]
          table {%
          0.1 0.00596058535600937
          0.075 0.00973483770480402
          0.05 0.00975868311643619
          0.025 0.00529486673506705
          0.01 0.00168587505107681
          0.001 0.000371620659272152
          };
          \addlegendentry{$ 4\cdot80$}
          \addplot [very thick,dashed, forestgreen4416044, forget plot]
          table {%
          0.1 0.00036115722171378
          0.075 0.00036115722171378
          0.05 0.00036115722171378
          0.025 0.00036115722171378
          0.01 0.00036115722171378
          0.001 0.00036115722171378
          };

      \nextgroupplot[
          legend cell align={left},
          legend style={font=\scriptsize,
            at={(0.03,0.97)},
            anchor=north west
          },
          xlabel={$\tau$},
          x dir=reverse,
          xmajorgrids,
          xtick = {0.001, 0.01, 0.1}, 
          xticklabels={$10^{-3}$,$10^{-2}$,$10^{-1}$ },
          xmin=0.00079, xmax=0.125,
          xtick style={color=black},
          xmode=log,
          y grid style={darkgray176},
          ylabel={element ratio},
          ymajorgrids,
          ymin=-0.044, ymax=1.028,
          ytick style={color=black},
          ytick = {0, 0.25, 0.5,0.75, 1}, 
          grid style = dashed
          ]
          \addplot [very thick, steelblue31119180, mark=*, mark size=2, mark options={solid}]
          table {%
          0.1 0.059375
          0.075 0.0666666666666667
          0.05 0.08125
          0.025 0.0958333333333333
          0.01 0.134375
          0.001 0.2765625
          };
          \addlegendentry{$4\cdot60$}
          \addplot [very thick, darkorange25512714, mark=*, mark size=2, mark options={solid}]
          table {%
          0.1 0.0697916666666667
          0.075 0.078125
          0.05 0.0927083333333333
          0.025 0.113020833333333
          0.01 0.152083333333333
          0.001 0.319270833333333
          };
          \addlegendentry{$4\cdot70$}
          \addplot [very thick, forestgreen4416044, mark=*, mark size=2, mark options={solid}]
          table {%
          0.1 0.0760416666666667
          0.075 0.0864583333333333
          0.05 0.0973958333333333
          0.025 0.1265625
          0.01 0.1703125
          0.001 0.341145833333333
          };
          \addlegendentry{$4\cdot80$}
  \end{groupplot}

\end{tikzpicture}
    \caption{Influence of the tolerance $\tau$ on the accuracy of the solution of the "ring" system. On the left, the mean displacement error (\Cref{eq:mean_error}) is plotted over $\tau$. The dashed lines are the mean displacement error without ECSW. 
    On the right, the increase of the ratio of evaluated elements is plotted.} 
    \label{fig:ring_uerrECSW}
\end{figure}
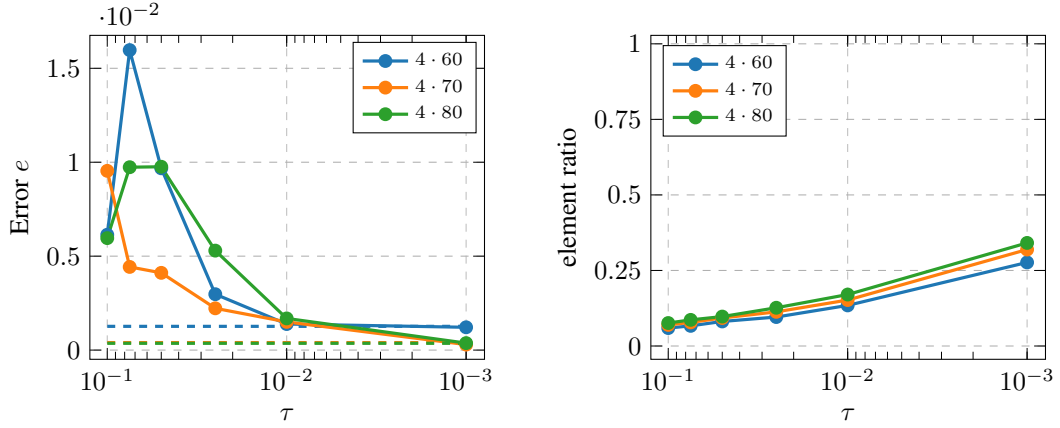

\newpage

\subsubsection{2D example}

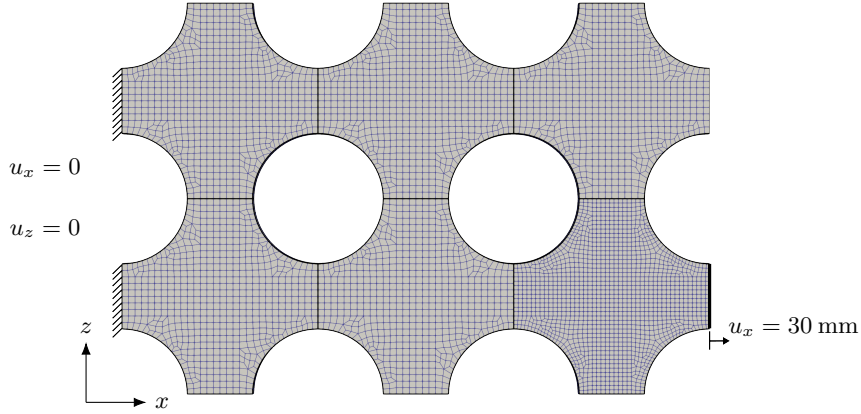
\begin{figure}[hptb]
    \centering
    {\small
    \begin{tikzpicture}
    \node[inner sep=0pt] (pic) at (0,0) {\includegraphics[width=0.5\textwidth]{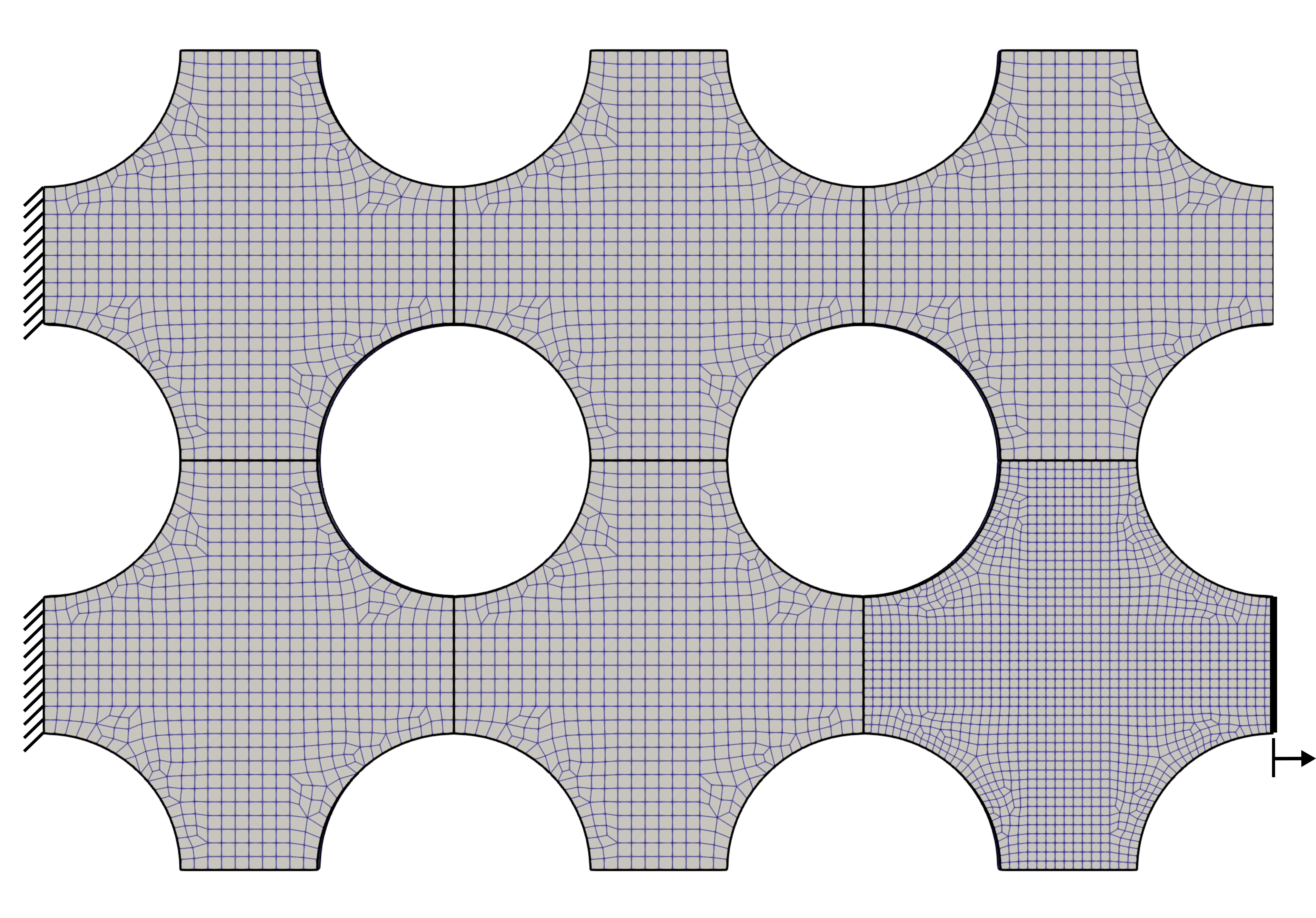} };
    \node[inner sep=0pt] at ($(pic.south) + (5,1.2)$) {$u_x = 30 \rm \: mm  $};
    \node[inner sep=0pt] at ($(pic.center) + (-4.9,0.4)$) {$u_x = 0  $};
    \node[inner sep=0pt] at ($(pic.center) + (-4.9,-0.4)$) {$u_z = 0  $};
    \draw[-{Latex}, thin] ($(pic.south west) + (-0.2,0.2)$) -- ++(0.8,0) node[right] {$x$}; 
    \draw[-{Latex}, thin] ($(pic.south west) + (-0.2,0.2)$) -- ++(0,0.8) node[above] {$z$};

    \end{tikzpicture}
    }
    \caption{Geometry, mesh and boundary conditions for the 2D example with $2\times 3$ substructures. Figure adapted from \citet{ritzert2025component}}. 
    \label{fig:geometry_23}
\end{figure}

The mesh and the boundary value problem are shown in \Cref{fig:geometry_23}. 
The structure is fixed on the left side in horizontal and vertical direction and a horizontal displacement of $u_x = 30 \rm mm$ is applied on the right side of the substructure on the bottom right. 
This substructure has a finer mesh because it undergoes the largest strains.

First, the influence of the centering of the snapshots is investigated. 
\Cref{fig:2_3_centered_normal} shows that the influence is very small. 
For all numbers of modes the accuracy is in a similar range. 
The reason is that the snapshots here do not contain rigid body motions because the boundary conditions in the sampling are applied directly to the component.
In the centering also volumetric deformations are removed because $\BPsi_{\rm RBM}$ contains volumetric deformation modes, as explained in \Cref{sec:RBM}. 
This leads to the small differences between the modes of the centered and normal snapshots.

\begin{figure} 
    \centering
\begin{tikzpicture}

\definecolor{crimson2143940}{RGB}{214,39,40}
\definecolor{darkgray176}{RGB}{176,176,176}
\definecolor{darkorange25512714}{RGB}{255,127,14}
\definecolor{forestgreen4416044}{RGB}{44,160,44}
\definecolor{lightgray204}{RGB}{204,204,204}
\definecolor{steelblue31119180}{RGB}{31,119,180}

\begin{axis}[
width = 7cm, 
legend cell align={left},
legend style={
  font=\small,
  at={(1.03,1.0)},
  anchor=north west
},
x grid style={dashed},
xlabel={$\bar u_x/ \mathrm{mm} $},
xmajorgrids,
xmin=2.4375, xmax=31.3125,
xtick style={color=black},
y grid style={dashed},
ylabel={$\| \bU_{\text{MOR}} - \bU_{\text{MOR}} \| / \|\bU_{\text{Ref}} \|$},
ymajorgrids,
ymin=5.12777817699482e-05, ymax=0.0139781657234349,
ytick style={color=black}
]
\addplot [very thick, steelblue31119180]
table {%
3.75 0.001366039129718
7.5 0.00212038007620671
11.25 0.00393339640137544
15 0.00614557251438161
18.75 0.00832921877192085
22.5 0.0102881938221786
26.25 0.0119595934188268
30 0.0133451253624501
};
\addlegendentry{$6\cdot30$ normal}
\addplot [very thick, darkorange25512714]
table {%
3.75 0.0010790004493952
7.5 0.00171356639066484
11.25 0.00304583734953483
15 0.00471224616750348
18.75 0.00641915022908539
22.5 0.00800254224900676
26.25 0.00939439105889065
30 0.0105811151350059
};
\addlegendentry{$6\cdot35$ normal}
\addplot [very thick, forestgreen4416044]
table {%
3.75 0.000684318142754719
7.5 0.00101464759919358
11.25 0.00156903021646078
15 0.00227023151851931
18.75 0.00301383598264894
22.5 0.00373866728458349
26.25 0.00441521996625818
30 0.00503167176007082
};
\addlegendentry{$6\cdot40$ normal}
\addplot [very thick, crimson2143940]
table {%
3.75 0.00059920942986875
7.5 0.000862684566622702
11.25 0.00119780773097256
15 0.00162790281704424
18.75 0.00210579485832023
22.5 0.00259070922095415
26.25 0.00305898675165407
30 0.00349862703133904
};
\addlegendentry{$6\cdot45$ normal}
\addplot [very thick,dashed, steelblue31119180]
table {%
3.75 0.000992828386826469
7.5 0.00220730697679971
11.25 0.00427132303337381
15 0.00650280991651905
18.75 0.00856675763990302
22.5 0.0103440077265889
26.25 0.0118168196696586
30 0.013010500548333
};
\addlegendentry{$6\cdot30$ centered}
\addplot [very thick,dashed, darkorange25512714]
table {%
3.75 0.000846186607559286
7.5 0.00158332736508382
11.25 0.00295839332181499
15 0.00456459264732519
18.75 0.00612451032696581
22.5 0.00751839890652299
26.25 0.00871088214926733
30 0.0097072410170906
};
\addlegendentry{$6\cdot35$ centered}
\addplot [very thick,dashed, forestgreen4416044]
table {%
3.75 0.000695287639711882
7.5 0.000971324447507726
11.25 0.00150558740240887
15 0.00221575692805569
18.75 0.00297015830166441
22.5 0.00370038609118258
26.25 0.00437691141709761
30 0.0049894727869159
};
\addlegendentry{$6\cdot40$ centered}
\addplot [very thick,dashed, crimson2143940]
table {%
3.75 0.000660409913706901
7.5 0.00087920494581201
11.25 0.00117526558438754
15 0.00159230601678858
18.75 0.00207344660795768
22.5 0.00256822492386585
26.25 0.00304783620995362
30 0.00349792583171359
};
\addlegendentry{$6\cdot45$ centered}
\end{axis}

\end{tikzpicture}
    \caption{Influence of the centering of the snapshots (see \Cref{sec:snapsPreprocessing}) on the accuracy of the solution for the $2\times3$ system. The displacement error is plotted over the applied displacement for different numbers of modes per substructure. } 
    \label{fig:2_3_centered_normal}
\end{figure}
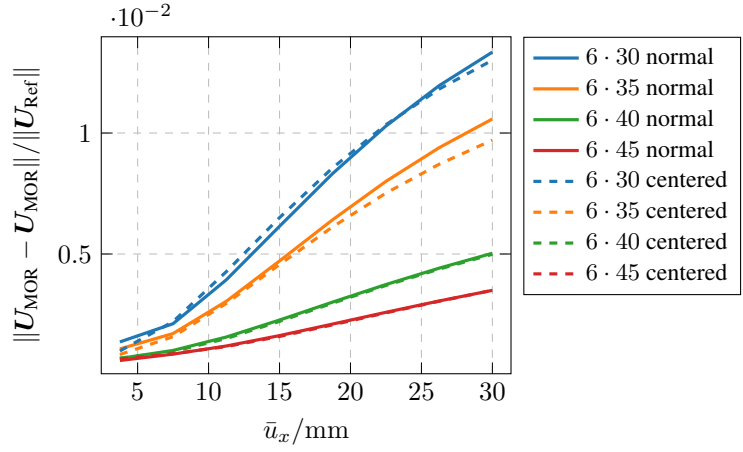

\begin{figure}[htbp]
    \centering
    \begin{tikzpicture}

  \definecolor{crimson2143940}{RGB}{214,39,40}
  \definecolor{darkgray176}{RGB}{176,176,176}
  \definecolor{darkorange25512714}{RGB}{255,127,14}
  \definecolor{forestgreen4416044}{RGB}{44,160,44}
  \definecolor{lightgray204}{RGB}{204,204,204}
  \definecolor{steelblue31119180}{RGB}{31,119,180}
  
  \begin{groupplot}[group style = {group name = group, group size = 2 by 1, horizontal sep = 2.2 cm}, width = 6.8 cm    ]  
      \nextgroupplot[
          legend cell align={left},
          legend style={font=\scriptsize },
          x dir=reverse,
          xlabel={$\tau $},
          xmajorgrids,
          xtick = {0.001, 0.01, 0.1}, 
          xticklabels={$10^{-3}$,$10^{-2}$,$10^{-1}$ },
          xmin=0.00079, xmax=0.125,
          xtick style={color=black},
          xmode=log,
          y grid style={darkgray176},
          ylabel={Error $ e  $},
          ymajorgrids,
          ymin=3.67544409775257e-05, ymax=0.0297482981294505,
          ytick style={color=black},
          grid style = dashed
          ]
          \addplot [very thick, steelblue31119180, mark=*, mark size=2, mark options={solid}]
          table {%
          0.1 0.0246025580336268
          0.05 0.00778566830988751
          0.01 0.0052056127559607
          0.005 0.00598717588506715
          0.001 0.00482793019776945
          };
          \addlegendentry{ $6\cdot35$}
          \addplot [very thick, dashed,forget plot, steelblue31119180]
          table {%
          0.1 0.00458097901227021
          0.05 0.00458097901227021
          0.01 0.00458097901227021
          0.005 0.00458097901227021
          0.001 0.00458097901227021
          };
          \addplot [very thick, darkorange25512714, mark=*, mark size=2, mark options={solid}]
          table {%
          0.1 0.0283977734163381
          0.05 0.00995468996020912
          0.01 0.00270627364177141
          0.005 0.00274132466262884
          0.001 0.00224429525326414
          };
          \addlegendentry{ $6\cdot40$}
          \addplot [very thick, dashed,forget plot, darkorange25512714]
          table {%
          0.1 0.00233368370989676
          0.05 0.00233368370989676
          0.01 0.00233368370989676
          0.005 0.00233368370989676
          0.001 0.00233368370989676
          };
          \addplot [very thick, forestgreen4416044, mark=*, mark size=2, mark options={solid}]
          table {%
          0.1 0.00861254530896225
          0.05 0.00902122671988441
          0.01 0.0027317657708717
          0.005 0.00159847167135138
          0.001 0.00138766468255383
          };
          \addlegendentry{ $6\cdot45$}
           \addplot [very thick, dashed,forget plot, forestgreen4416044]
          table {%
          0.1 0.00138727915408993
          0.05 0.00138727915408993
          0.01 0.00138727915408993
          0.005 0.00138727915408993
          0.001 0.00138727915408993
          };

      \nextgroupplot[
          legend cell align={left},
          legend style={font=\scriptsize,
            at={(0.03,0.97)},
            anchor=north west
          },
          xlabel={$\tau$},
          x dir=reverse,
          xmajorgrids,
          xtick = {0.001, 0.01, 0.1}, 
          xticklabels={$10^{-3}$,$10^{-2}$,$10^{-1}$ },
          xmin=0.00079, xmax=0.125,
          xtick style={color=black},
          xmode=log,
          y grid style={darkgray176},
          ylabel={element ratio},
          ymajorgrids,
          ymin=-0.044, ymax=1.028,
          ytick style={color=black},
          ytick = {0, 0.25, 0.5,0.75, 1}, 
          grid style = dashed
          ]
          \addplot [very thick, steelblue31119180, mark=*, mark size=2, mark options={solid}]
          table {%
          0.1 0.0553702468312208
          0.05 0.0787191460973983
          0.01 0.13030909495219
          0.005 0.155881698910385
          0.001 0.241049588614632
          };
          \addlegendentry{$6\cdot35$}               
          \addplot [very thick, darkorange25512714, mark=*, mark size=2, mark options={solid}]
          table {%
          0.1 0.0618189904380698
          0.05 0.0851678897042473
          0.01 0.145430286857905
          0.005 0.176562152546142
          0.001 0.275294640871692
          };
          \addlegendentry{$6\cdot40$}
          \addplot [very thick, forestgreen4416044, mark=*, mark size=2, mark options={solid}]
          table {%
          0.1 0.079608627974205
          0.05 0.102290415832777
          0.01 0.159439626417612
          0.005 0.190126751167445
          0.001 0.29997776295308
          };
          \addlegendentry{$6\cdot45$}
  \end{groupplot}

\end{tikzpicture}
    \caption{Influence of the tolerance $\tau$ on the accuracy of the solution of the $2\times3$ system. On the left, the mean displacement error is plotted over $\tau$. The dashed lines are the mean displacement error without ECSW. 
    On the right, the increase of the ratio of evaluated elements is plotted.} 
    \label{fig:kreuz3x2_uerrECSW}
\end{figure}

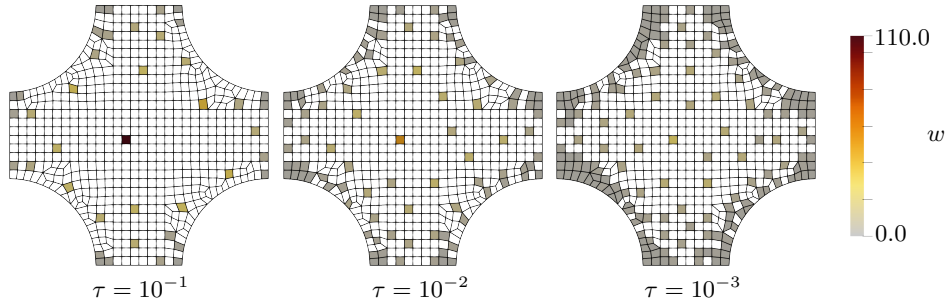
\begin{figure}[hptb]
    \centering
    {\small
    \begin{tikzpicture}
    \node[inner sep=0pt] (pic) at (0,0) {\includegraphics[width=0.7\textwidth]{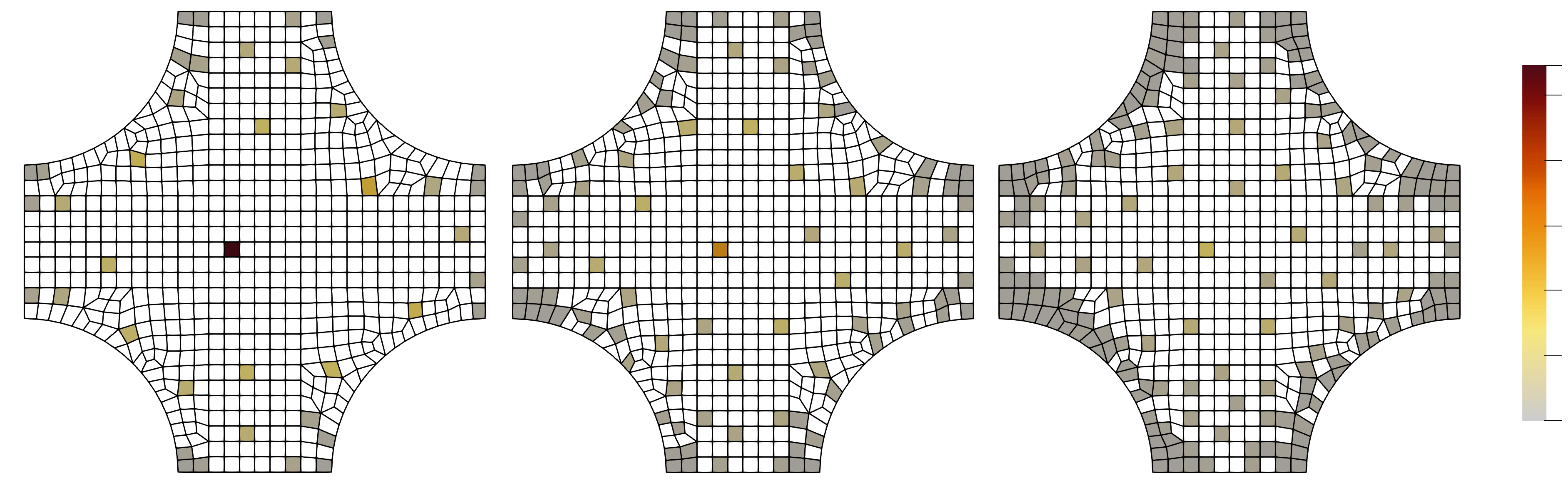} };
    \node[inner sep=0pt] at ($(pic.south) + (-3.9,-0.20)$) {$ \tau= 10^{-1} $};
    \node[inner sep=0pt] at ($(pic.south) + (-0.2,-0.20)$) {$ \tau= 10^{-2} $};
    \node[inner sep=0pt] at ($(pic.south) + (3.4,-0.2)$) {$ \tau= 10^{-3} $};
    \node[inner sep=0pt] at ($(pic.east) + (0.35,1.3)$) {$ 110.0 $};
    \node[inner sep=0pt] at ($(pic.east) + (0.2,-1.3)$) {$ 0.0 $};
    \node[inner sep=0pt] at ($(pic.east) + (0.8,0)$) {$ w $};
    \end{tikzpicture}
    }
    \caption{Visualization of the ECSW weights for the substructure used in the $2\times3$ system. The colorbar shows the weights of the elements. The three subfigures show the weights for different tolerances $\tau$.} 
    \label{fig:kreuz_ecsw_elements}
\end{figure}

\Cref{fig:kreuz3x2_uerrECSW} shows that the errors converge to the errors of the POD solution if the tolerance $\tau$ is decreased. 
The dashed lines are the mean displacement error over all time steps of the reduced system without ECSW.  
The right plot shows the growing amount of evaluated elements for a decreasing $\tau$.   
For sufficiently converged results, a tolerance of $\tau=10^{-2}$ is necessary, but with a tolerance of $\tau=0.05$ errors below $1 \%$ can already be achieved. 
For $\tau=10^{-2}$, $15.9 \%$ of the elements need to be evaluated when 35 POD modes are used per substructure. 
In \Cref{fig:kreuz_ecsw_elements} the selected elements and their weights are visualized for three different values of $\tau$. 
Similar to the 3D examples, the selected elements are located predominantly at the interfaces of the substructure as $\tau $ decreases.  
These elements at the boundary have comparatively small weights. 
The elements with the highest weights are located in the center of the substructure.

\subsection{Larger examples} 
In this section, the method is applied to structures composed of many substructures. 
First, we will show the predictive capabilities for a quasi static case and then for a finite-strain dynamic example.  

\subsubsection{Quasi static simulations}

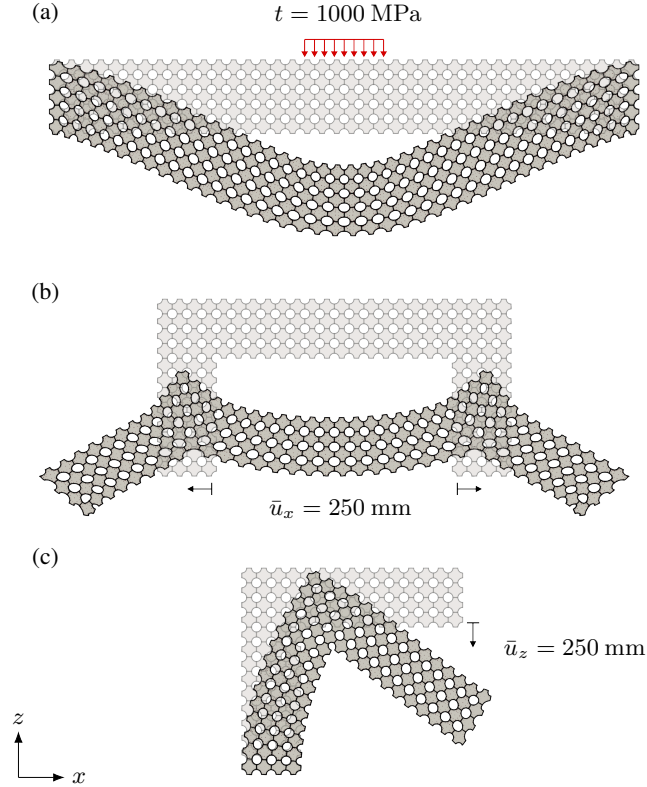
\begin{figure}[hptb]
    \centering
    {\small
    \begin{tikzpicture}
    \node[inner sep=0pt] (pic) at (0,0) {\includegraphics[width=0.5\textwidth]{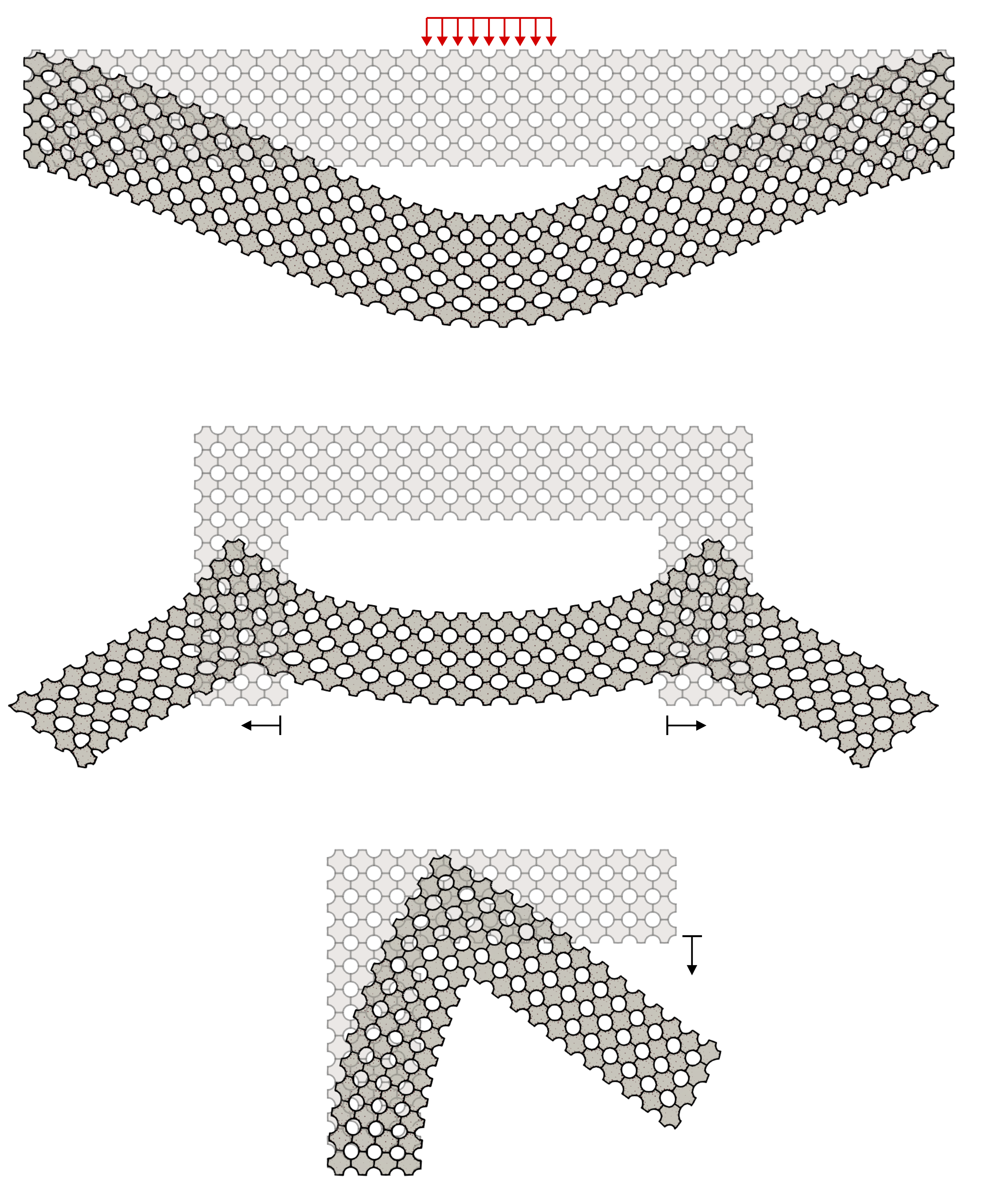} };
    \node[inner sep=0pt] at ($(pic.south) + (3,1.9)$) {$ \bar u_z = 250 \rm \: mm  $};
    \node[inner sep=0pt] at ($(pic.center) + (-0.1,-1.3)$) {$\bar u_x = 250 \rm \: mm  $};
    \node[inner sep=0pt] at ($(pic.north) + (-0,0.2)$) {$t = 1000 \: \rm MPa  $};
    \node[inner sep=0pt] at ($(pic.north) + (-4,0.2)$) {(a)};
    \node[inner sep=0pt] at ($(pic.north) + (-4,-3.5)$) {(b)};
    \node[inner sep=0pt] at ($(pic.north) + (-4,-7)$) {(c)};
    \draw[-{Latex}, very thin] ($(pic.south west) + (-0.2,0.2)$) -- ++(0.6,0) node[right] {$x$}; 
    \draw[-{Latex}, very thin] ($(pic.south west) + (-0.2,0.2)$) -- ++(0,0.6) node[above] {$z$};
    \end{tikzpicture}
    }
    \caption{Sketch of three different assemblies and their corresponding boundary conditions in the undeformed and deformed configuration.  } 
    \label{fig:KreuzStructures}
\end{figure}

In \Cref{fig:KreuzStructures} we show three different assemblies under different loading conditions. 
For all systems, we show the deformed configuration with the substructure boundaries. 
The undeformed configuration is displayed in the background.
We consider a compressible Neo-Hookean material model with $\lambda = 3727 \rm \: MPa$ and $\mu = 8696 \rm \: MPa$. 
These parameters are different from the parameters used in the snapshot sampling (cf. \Cref{sec:snapshot_sampling}).  
We use 35 POD modes per substructure and a tolerance of $\tau = 0.05$, leading to 59 elements per substructure that need to be evaluated. 
This number is the same for the coarse and the fine mesh used in the three systems. 

The results of the simulations are summarized in \Cref{tab:KreuzStructuresResults}. 
The differences in the computational time are due to the different amount of fine and coarse meshes in the three systems. 
When the mesh is refined, the speed-up of the ECSW method increases because the number of evaluated elements is independent of the mesh. 
For the system (c) we can achieve the highest speed-up of $t_{\rm ref}/ t_{\rm ECSW} = 12.536$ because 32 of the 100 substructures are discretized using the finer mesh.




\begin{table}[h] 
\centering 
\caption{Summary of the results for the three different assemblies from \Cref{fig:KreuzStructures}.}
\begin{tabular}{@{}lcccccc@{}}
\toprule
  System & DOFs  & reduced DOFs  & element ratio & displacement error & speed-up & time ratio \\ 
    &  $n$   &  $m$ & $\frac{\|\bar\cE \| }{ \|\cE \|}$ & $\frac{\| \bU_{MOR} - \bU_{\rm ref} \|}{\|\bU_{\rm ref}\|} $ &$\frac{t_{\rm ref}}{t_{\rm ECSW}}$ & $\frac{t_{\rm ECSW}}{t_{\rm ref}}$ \\ 
  \midrule
  (a) & 434490 & 3500     & 0.089  & 0.0094 & 9.667 & 0.103 \\ 
  (b) & 328800 & 2800     & 0.095  & 0.0086 & 8.41 & 0.119 \\ 
  (c) & 561336 & 3500     & 0.068  & 0.0051 & 12.536 & 0.080 \\ 
 \bottomrule
\end{tabular}
\label{tab:KreuzStructuresResults}
\end{table}

It can be seen that the solution of all three systems could be predicted with an error below 1 \%. 
When finer meshes were used, the speed-ups became higher because the number of evaluated elements in the reduced mesh is almost completely independent of the fidelity of the mesh. 
In general, we observed that the number of ECSW elements depends mostly on the number of modes and the tolerance $\tau$. 
Mesh refinement only leads to small differences in the number of elements that need to be evaluated.


\subsubsection{Finite-strain dynamic example }\label{sec:dynExample}
With this numerical example, we want to show that the method can also be applied to nonlinear dynamic substructuring problems. 
Additionally, we show that we can predict finite-strain hyper-viscoelastic material behavior with POD modes and ECSW weights and elements that were computed with elastic material behavior.  

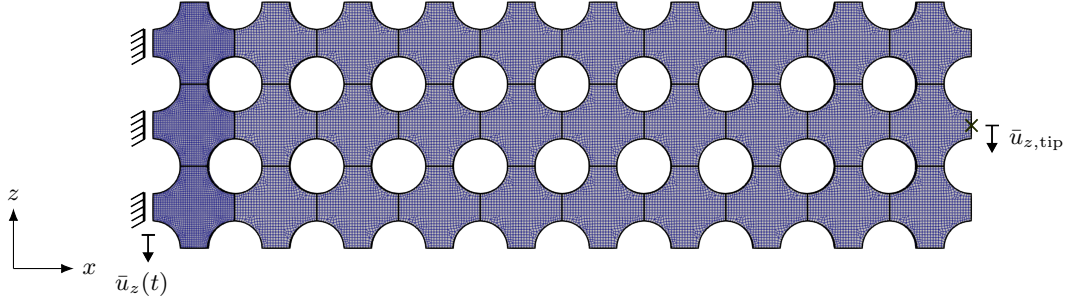
\begin{figure}[hptb]
    \centering
    {\small
    \begin{tikzpicture}
    \node[inner sep=0pt] (pic) at (0,0) {\includegraphics[width=0.7\textwidth]{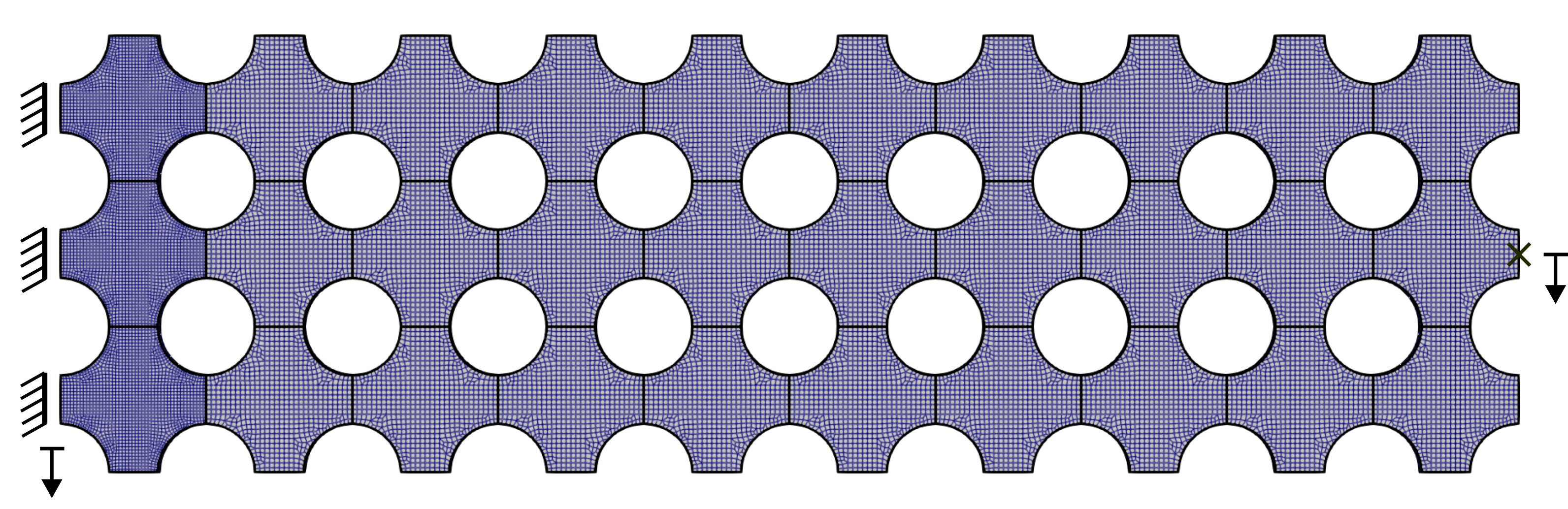} };
    \node[inner sep=0pt] at ($(pic.west) + (0.3,-2.1)$) {$ \bar u_z(t)   $};
    \node[inner sep=0pt] at ($(pic.east) + (0.5,-0.2)$) {$ \bar u_{z,\rm tip}   $};
    \draw[-{Latex}, thin] ($(pic.south west) + (-1.4,-0.0)$) -- ++(0.8,0) node[right] {$x$}; 
    \draw[-{Latex}, thin] ($(pic.south west) + (-1.4,-0.0)$) -- ++(0,0.8) node[above] {$z$};
    \end{tikzpicture}
    }
    \caption{Sketch of the mesh and the boundary value problem of the forced vibration example. The time-dependent $z$-displacements on the left side are given by the function \Cref{eq:inputsignal}. On the right side we measure the $z$-displacements at the tip. } 
    \label{fig:bvp_vibration}
\end{figure}

\begin{figure}
    \centering
    \input{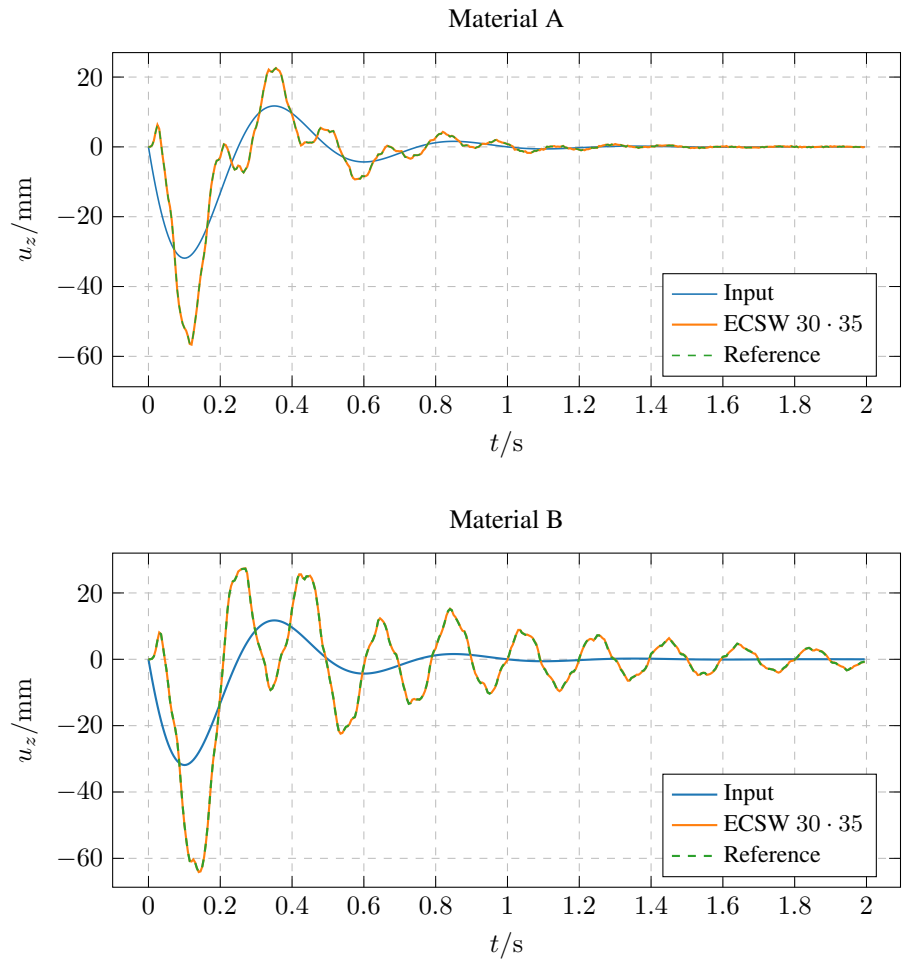}
    \caption{Plots of the tip displacement $ u_z^{\rm tip} $ over time for two different sets of material parameters. } 
    \label{fig:beam_dyn}
\end{figure}

\Cref{fig:bvp_vibration} shows the boundary value problem of the forced vibration simulation. 
On the left side, we apply a time-dependent displacement in $z$-direction and fix the displacement in $x$-direction.  
On the right side, we measure the $z$-displacement at the tip. 
The input function is the damped sine wave 
\begin{equation}\label{eq:inputsignal} 
\bar u_z(t) = -  \frac{3}{2} e^{-5\, t} \, \sin(5 \pi \, t) \, 50  \: \: \rm{mm}. 
\end{equation}
This function is also shown in \Cref{fig:beam_dyn}.

The general hyper-viscoelastic material model used here, is derived in \Cref{sec:viscoModel}. 
This general model is specified by choosing suitable functions for the Helmholtz free energy $\psi(\bC,\bar\bC_e)$ and the dissipation potential $g(\bar\bZ) $.
The Helmholtz free energy consists of an equilibrium and a non-equilibrium part 
\begin{equation} 
    \psi = \psi_{eq}(\bC) + \psi_{neq}(\bar\bC_e),
\end{equation}
where both energies are of Neo-Hookean type
\begin{align}
    \psi_{eq}(\bC) &= \frac{\mu_{eq}}{2} \left( \tr(\bC) -3 - \ln(\det(\bC)) \right) + \frac{\lambda_{eq}}{4} \left( \det(\bC) -1 - \ln(\det(\bC)) \right) \\ 
    \psi_{neq}(\bar \bC_e) &= \frac{\mu_{neq}}{2} \left( \tr(\bar \bC_e) -3 - \ln(\det(\bar \bC_e)) \right) + \frac{\lambda_{neq}}{4} \left( \det(\bar\bC_e) -1 - \ln(\det(\bar\bC_e)) \right).
\end{align}
The dissipation potential is chosen as 
\begin{equation} 
    g(\bar\bZ) = \frac{1}{2\mu_{neq}} \tr \left( \operatorname{dev}(\bar\bZ)^2 \right) + \frac{1}{9\kappa} \tr(\bar\bZ)^2,
\end{equation}
with $\kappa=\lambda_{neq} + 2/3 \mu_{neq}$.

To investigate if the POD bases shown in \Cref{sec:samplingKreuz} can be used for different material behavior, we consider two different sets of material parameters
These different parameter sets result in different damping behavior. 
The parameters are listed in \Cref{tab:MatPars}. 
\begin{table}[h] 
\centering 
\caption{Material parameters for the finite strain hyper-viscoelastic material model.} 
\begin{tabular}{@{}llll@{}}
\toprule
  &Material A  & Material B & unit   \\ \midrule
  $\lambda_{\rm eq}$ & 932    &  932 & MPa \\ 
  $\mu_{\rm eq}$ & 2174    &  2174 & MPa \\ 
  $\lambda_{\rm neq}$ & 3726   &  1864 & MPa \\ 
  $\mu_{\rm neq}$ & 8696    &  4348 & MPa \\ 
  $\eta$ & 4 & 2 & 1/s \\
 \bottomrule
\end{tabular}
\label{tab:MatPars}
\end{table}

\paragraph{Evaluation of the reduced simulation} 
The system is solved with 40 modes per substructure and an ECSW tolerance of $\tau=0.05$.  
The ratio of the unreduced degrees of freedom $n$ and the reduced degrees of freedom $m$  is $n/m = 137394/1200 = 114.5 $. 
For the solution we evaluate 59 elements per substructure for both the fine and coarse discretized substructures. 
In this system 3 of the 30 substructures have a finer mesh.
The element ratio of reduced mesh to the full mesh is  $\|\bar\cE \| / \|\cE \| = 0.084$.

For the analysis of the accuracy we use two different error measures. 
The first error measure is a cross-correlation error of the tip displacement 
\begin{equation} 
    e = \frac{u_z^{\rm tip, MOR} \cdot u_z^{\rm tip, Ref} }{u_z^{\rm tip, Ref} \cdot u_z^{\rm tip, Ref} } - 1.
\end{equation} 
This error can capture if the shape of the tip displacement is predicted correctly. 
Slight phase shifts of the tip displacement can lead to large errors in the mean squared error, but the cross-correlation error is not affected by this.
Additionally, we compute the mean squared error over all time steps for the full field 
\begin{equation} 
    \mathrm{MSE} =  \frac{1}{n_{\rm steps}} \sum_{t=1}^{n_{ \rm steps}} \left( \bU_t^{\rm MOR} - \bU_t^{\rm Ref}\right)^2.
\end{equation}

For material A, the cross-correlation error is $e= 1.8 \, 10^{-4}$ and the mean squared error of the full displacement field is $\mathrm{MSE} = 5.1 \, 10^{-4}$.  
For material B the errors are roughly twice as large because the deformations are also larger. Here, the cross-correlation error is  $e= 2.9 \, 10^{-4}$ and the mean squared error of the full displacement field is $\mathrm{MSE} = 2.5 \, 10^{-3}$. 
In general the errors are very small and the tip displacement can be predicted with high accuracy. 
In \Cref{fig:beam_dyn} the tip displacement over time is plotted for both materials.
The reference solution is plotted as a dashed line and the solution of the reduced system is plotted as a solid line. 
Visually, the solutions are indistinguishable.

\section{Conclusions and outlook}\label{sec:conclusion} 

We developed a component-wise hyperreduction method for nonlinear solid mechanics problems, where the hyperreduced components can be assembled into different systems.  
The component-wise hyperreduction is based on POD and ECSW on the component level and the substructures are connected by mortar mesh tying. 

The projection matrices are computed offline using single component simulations with parametrized boundary conditions on possible interfaces. 
The snapshots data matrix contains twelve additional modes to model finite rotations and translations. 
The ECSW weights and elements are computed from the same snapshots in the offline phase using the full substructure projection matrix. 
In the online phase, the POD basis, the weights, and the ECSW elements are reused for different assemblies and boundary conditions.
For the mesh tying, the slave side interface degrees of freedom are projected onto the POD space of the master side.  

In the numerical examples, we could show that the substructure POD bases and ECSW meshes can be used for different assemblies and boundary conditions. 
They are robust under changes in material parameters and can be used also for finite strain hyper-viscoelasticity, even though the snapshots were generated with a purely hyperelastic Neo-Hookean material. 
We showed that finite rotations and translations can be described by adding three translation and nine rotation modes to the snapshot matrix.  
We showed how the rotation modes can be derived from the general description of finite rotations and how they can be used to improve the accuracy of the reduced order model. 
The advantage of this approach is its simplicity. With only a few additional modes, the reduced order model can be used for large rotations and translations.
The results show that the component-wise hyperreduction can significantly reduce the computational cost, while maintaining good accuracy. 
The hyperreduced simulations converge toward the POD solution with increasing number of ECSW elements. 
The key result is not only that ECSW accelerates a reduced model, but that the ECSW-selected component meshes remain reusable after assembly. 
This suggests that hyperreduced components can serve as transferable simulation building blocks for modular nonlinear solid mechanics.

In future work, we will investigate more efficient hyperreduction methods for the substructures and apply the method to more complex inelastic problems. 
The assembly of the hyperreduced components is currently very intrusive and requires access to the solver. 
We plan work on making the method non-intrusive and implement it in a commercial solver.

\section{Declaration of competing interest} 
The authors declare that they have no known competing financial interests or personal relationships that could have appeared
to influence the work reported in this paper.

\section{Data availability} 
Data will be made available on request. 


\section{Acknowledgements}  
The authors gratefully acknowledge the funding granted by the German Research Foundation (DFG). 
The results presented here were developed within the subproject A01 of the Transregional Collaborative Research Center (CRC) Transregio (TRR) 280 with project number 444299690. 
Furthermore, T. Brepols, J. Kehls, and S. Reese gratefully acknowledge the funding that was granted within the subproject B05 "Coupling of intrusive and non-intrusive locally decomposed model order reduction techniques for rapid simulations of road systems" of the DFG CRC/TRR 339 with the project number 496338782, that was strongly involved in the origin of the paper.


\appendix
\section{Appendix}
\label{sec:appendix}

\subsection{Finite strain hyper-viscoelastic material model}\label{sec:viscoModel}
For the derivation of the hyper-viscoelastic material model, we follow the method described by \citet{holthusen_inelastic_2023} and define the inelastic quantities in the co-rotated intermediate configuration. 
It is derived from a Helmholtz free energy depending on the total strain and the elastic strain and an evolution equation describing the viscous strains. 

The kinematics of the hyper-viscoelastic material are described by the multiplicative split of the deformation gradient $\bF = \bF_e \bF_i$ into elastic $\bF_e$ and inelastic parts $\bF_i$.
The two tensors can be split by a polar decomposition $\bF_e = \bV_e \bR_e, \:  \bF_i = \bR_i \bU_i$ into non-unique rotational parts $\bR_e$,$\bR_i$ and unique stretch tensors $\bV_e$, $\bU_i$. 
The material model is formulated using uniquely defined strain tensors in the reference configuration and in the co-rotated intermediate configuration. 
In the reference configuration we use the right Cauchy-Green strain $\bC = \bF^T \bF$ and the inelastic Cauchy-Green strain, which is defined as $\bC_i = \bU_i^2$. 
The co-rotated elastic Cauchy-Green strain can be computed by a partial push forward $\bar \bC_e = \bU_i^{-1} \bC \bU_i^{-1}$ by the inelastic stretch tensor $\bU_i$. 
All of these strain tensors are not affected by the rotational non-uniqueness. 
Based on this stretch, we can further define the co-rotated rate of deformation tensor
\begin{equation}\label{eq:D_i}
\bar \bD_i = \operatorname{sym}(\dot \bU_i \bU_i^{-1}) =\frac{1}{2}\bU_i^{-1} \dot\bC_i  \bU_i^{-1}. 
\end{equation}

The stress and the thermodynamic driving force are derived from the Helmholtz free energy $\psi= \psi(\bC,\bar\bC_e)$, depending on the right Cauchy-Green strain $\bC$ and the co-rotated elastic Cauchy-Green strain $\bar \bC_e$.  
The second Piola Kirchhoff stress is derived from the Clausius-Duhem inequality by using the standard Coleman-Noll procedure, leading to
\begin{equation} 
    \bS = 2 \frac{\partial \psi}{\partial \bC} + \bU_i^{-1} \frac{\partial \psi}{\partial \bar \bC_e}\bU_i^{-1}. 
\end{equation} 
This stress definition leads to the remaining dissipation inequality $\bar \bZ \cdot \bar \bD_i \ge 0$, where the thermodynamic driving force $\bar \bZ$ is defined as 
\begin{equation} 
    \bar \bZ = 2 \bar \bC_e \frac{\partial \psi}{\partial \bar\bC_e}.
\end{equation}
The remaining dissipation inequality can be fulfilled if the evolution equation of $\bar \bD_i$ is derived from a non-negative and convex potential $g(\bar \bZ)$ that is assumed to be an isotropic function of $\bar \bZ$. 
Using \Cref{eq:D_i}, we can write the evolution equation in terms of the inealstic right Cauchy-Green strain as 
\begin{equation} 
    \dot \bC_i = 2 \eta \bU_i \frac{\partial g(\bar \bZ) }{\partial \bar \bZ} \bU_i. 
\end{equation}
Here, $\eta$ is a material parameter controlling the rate of dissipation.
In a time discretized setting on the interval $t\in [t_n, t_{n+1}]$ with $\Delta t = t_{n+1} -t_n$, we solve the evolution equation by an exponential map algorithm (see e.g. \citep{dettmer2004theoretical,vladimirov2008modelling, christ2009finite}).
After some derivation steps shown in the work \citet{holthusen_inelastic_2023}, one arrives at the residual equation that is solved at each integration point
\begin{equation} 
    r(\bU_i) = \bU_i \exp\left(-2\Delta t \, \eta \frac{\partial g(\bar\bZ)}{\partial\bar \bZ} \right) \bU_i - \bU_{i\, n}^2  = 0 , 
\end{equation}
where we dropped the index $n+1$. 
It can be seen that the internal variable in this material model is the inelastic stretch $\bU_i$. 
This residual equation is solved by a Newton-Raphson procedure. 




\bibliographystyle{plainnat}
\bibliography{literature}

\end{document}